\documentclass[sigconf,edbt]{acmart-edbt2019}

\usepackage{array}
\usepackage{balance}
\usepackage{graphicx}
\usepackage{multirow} 

\theoremstyle{definition}

\setcopyright{rightsretained}

\acmDOI{}

\acmISBN{XXX-X-XXXXX-XXX-X}

\acmConference[EDBT 2019]{22nd International Conference on Extending Database Technology (EDBT)}{March 26-29, 2019}{Lisbon, Portugal} 
\acmYear{2019}

\begin{document}

\title{Comparative Analysis of Content-based Personalized \\Microblog Recommendations
\LARGE{[Experiments and Analysis]}}
% \subtitle{Experiments and Analyses Track}
%\vspace{-25pt}

\author{Efi Karra Taniskidou}
\affiliation{%
  \institution{University of California, Irvine}
}
\email{ekarrata@uci.edu}

\author{George Papadakis}
%\authornote{The secretary disavows any knowledge of this author's actions.}
\affiliation{%
  \institution{University of Athens}
}
\email{gpapadis@di.uoa.gr}

\author{George Giannakopoulos}
\affiliation{%
  \institution{NCSR Demokritos}
}
\email{ggianna@iit.demokritos.gr}

\author{Manolis Koubarakis}
\affiliation{%
  \institution{University of Athens}
}
\email{koubarak@di.uoa.gr}

% \renewcommand{\shortauthors}{B. Trovato et al.}
\renewcommand{\shortauthors}{}

\begin{abstract}
Microblogging platforms constitute a popular means of real-time communication
and information sharing. They involve such a large volume of user-generated
content that their users suffer from an information deluge. To address it,
numerous recommendation methods have been proposed to organize the posts a user
receives according to her interests. The content-based methods typically build a text-based
model for every individual user to capture her tastes and then rank the posts
in her timeline according to their similarity with that model.
Even though content-based methods have attracted lots of interest in the data management
community, there is no comprehensive evaluation of the main factors that affect their performance. These are:  \emph{(i)} the representation model that converts an unstructured text into a structured representation that elucidates its characteristics, \emph{(ii)}
the source of the microblog posts that compose the user models, and \emph{(iii)} the type of user's posting activity. To cover this gap, we systematically examine
the performance of 9 state-of-the-art representation models in combination with 13 representation sources and 3 user types over a large, real dataset from Twitter comprising 60 users. 
We also consider a wide range of 223 plausible configurations for the representation models in order to 
assess their robustness with respect to their internal parameters. To facilitate the interpretation of our experimental results, we introduce a novel taxonomy of representation models.
Our analysis provides novel insights into 
%the performance and functionality of 
the main factors determining the performance of content-based recommendation in microblogs.
\end{abstract}

\maketitle

%\vspace{-5pt}
\section{Introduction}

Microblogging platforms enable the instant communication and interaction
between people all over the world. They allow their users to post messages in
real-time, often carelessly and ungrammatically, through any electronic device,
be it a mobile phone or a personal computer. They also allow for explicit
connections between users so as to facilitate the dissemination and consumption
of information. These characteristics led to the explosive growth of
platforms like 
% microblogging services in recent years, with the most popular ones being
Twitter (www.twitter.com), Plurk (www.plurk.com), Sina Weibo (www.weibo.com) and Tencent Weibo (http://t.qq.com).

Their popularity has led to an information deluge: the number of
messages that are transmitted on a daily basis on Twitter alone has jumped from
35 million tweets in 2010 to over 500 million in
2017 \cite{DBLP:conf/edbt/GrossettiCMT18}. Inevitably, their users are constantly overwhelmed with information.
As we also show in our experiments, this situation cannot be ameliorated by presenting the new messages in chronological order; the relatedness with
users' interests is typically more important than the recency of a post. Equally ineffective is the list of trending topics, where the same messages are presented to all users,
irrespective of their personal interests.

A more principled solution to information deluge is
offered by \textit{Personalized Microblog Recommendation} (PMR). Its goal is to
capture users' preferences so as to direct their attention to the messages that
better match their personal interests.
%among a large number of different options. 
A plethora of works 
%in the literature 
actually focuses on \textit{Content-based PMR} \cite{balabanovic1997fab,chen2009make,chen2010short,hannon2010recommending,kywe2012recommending,lu2012twitter,mooney2000content}, which typically operates as follows: 
first, it builds a document model for every individual post in the training set by
extracting features from its textual content. Then, it constructs a user model by assembling the document models that capture the user's
preferences. Subsequently, it compares the user model to the models of
recommendation candidates (documents) with a similarity measure. The resulting
similarity scores are used to rank all candidates in descending order,
from the highest to the lowest score, thus placing the most relevant
ones at the top positions. Finally, the ranked list is presented to the user.

Content-based PMR is a popular problem that has
attracted a lot of attention in the data management
community \cite{abel2011analyzing,chen2010short,DBLP:conf/edbt/GrossettiCMT18,DBLP:journals/pvldb/GuptaSGGZLL14,hannon2010recommending,DBLP:journals/pvldb/SharmaJBLL16}.
However, the experimental results presented in the plethora of relevant works are not directly comparable, due to the different configurations that are used for several important, yet overlooked parameters. 

The core parameter is the \textit{representation model} that is used for converting a (set of) unstructured texts into
structured representations that reveal their characteristics. The available options range from traditional vector space models \cite{DBLP:conf/webi/ZhaoT14, DBLP:conf/icdm/KimS11} to topic models \cite{ramage2010characterizing,DBLP:conf/icdm/KimS11}. Also crucial is the \textit{representation source}, i.e., the source of the microblog posts that compose user models. Common choices include the user's tweets \cite{DBLP:journals/tkde/JiangCWZY14} together with their retweets \cite{chen2012collaborative, DBLP:conf/wsdm/FengW13, DBLP:journals/tist/ShenWZHYSG13, lu2012twitter} as well as the posts of followers \cite{hannon2010recommending,ramage2010characterizing,DBLP:conf/webi/ZhaoT14} and followees \cite{DBLP:conf/icdm/KimS11,chen2010short,hannon2010recommending}. Another decisive factor is the posting activity of a user, i.e., whether she is an information producer or seeker \cite{armentano2011topology,java2007we}. Other parameters include the novel challenges posed by the short, noisy, multilingual content of microblogs as well as the external information that enriches their textual content, e.g., concepts extracted from
Wikipedia \cite{lu2012twitter} or
the content of a Web page, whose URL is mentioned in a post
\cite{abel2011analyzing}.

Despite their significance,
little effort has been allocated on assessing the impact of these parameters on Content-based PMR. 
%relative performance in content recommendation. 
To cover this gap, we perform a thorough experimental analysis
that investigates the following questions: 
\textit{Which representation model is the most effective for recommending short, noisy, multilingual
microblog posts? Which is the most efficient one? 
% Which is the best configuration for each representation model? 
How robust is the
performance of each model with respect to its configuration? Which
representation source yields the best performance? How does the behavior of
individual users affect the performance of Content-based MPR?} We leave the investigation of external information as a future work, due to the high diversity of proposed approaches, which range from language-specific word embeddings like Glove \cite{pennington2014glove} to self-reported profile information \cite{DBLP:conf/kdd/El-AriniPHGA12}.
%Answers to these questions will facilitate practitioners and
%researchers to effectively use and extend the state-of-the-art representation
%models.

% \textbf{Representation Sources in Twitter.} Messages on Twitter are called
% \textit{tweets} or \textit{statuses} and are limited to contain up to 140
% characters. We refer to the messages posted by a specific user as her
% \textit{outgoing tweets}. Users choose to receive statuses of other accounts by
% \textit{following} them; we call the tweets a user receives from her followees
% as her \textit{incoming tweets}. Users can also retweet their followees’
% statuses to share interesting information with their own followers.

To investigate the above questions, we focus on Twitter, the most popular
microblogging service worldwide, with over 335 million active
users per
month.\footnote{\url{https://en.wikipedia.org/wiki/Twitter}, last accessed on 14 January 2019.} We begin with a categorization of the representation
sources and the users it involves, based on its special social graph: every user $u_1$ is allowed to
unilaterally follow another user $u_2$, with $u_1$ being a \textit{follower} of
$u_2$, and $u_2$ a \textit{followee} for $u_1$; if $u_2$ follows back $u_1$, the
two users are \textit{reciprocally connected}. Then, we list the novel
challenges posed by the short, noisy, user-generated tweets in comparison with the
long and curated content of traditional documents. We also introduce a taxonomy
of representation models that provides insights into their endogenous
characteristics. Based on it, we briefly present nine
state-of-the-art representation models and apply them to a dataset of 60
real Twitter users (partitioned into three different categories) in combination with
223 parameter configurations, three user types and 13 representation sources. Finally, we discuss the
experimental outcomes in detail, interpreting the impact of every parameter on the performance of~Content-based~PMR. 
%relative performance of the
%models with the help of our taxonomy.

In short, we make the following contributions:

$\bullet$ We perform the first systematic study for content recommendation
in microblogging platforms, covering 9 representation models, 13 representation
sources and 3 user types. We have publicly released our code along with
guidelines for our datasets\footnote{See \url{https://github.com/efikarra/text-models-twitter} for more details.}.

$\bullet$ We organize the main representation models according to their
functionality in a novel taxonomy with three main categories and two subcategories.
In this way, we facilitate the understanding of our experimental results, given that every
(sub-)category exhibits different behavior.

$\bullet$ We examine numerous configurations for every representation
model, assessing their relative effectiveness, robustness and time efficiency. Our conclusions
% We also identify the best configuration for every  model, 
facilitate their fine-tuning and use in real recommender systems.
% applications.
%  not only on Twitter, but also on other
% micro-blogging services with similar characteristics.
%  by researchers to select the most effective or robust
% representation strategy or to fine-tune a model they already use for recommendations,

The rest of the paper is structured as follows: Section~\ref{sec:preliminaries} provides background
knowledge on Twitter and formally defines the recommendation task we are
tackling in this work.
In Section~\ref{sec:representationModels}, we present our taxonomy of representation models
and  
%and in Section~\ref{sec:representationModels}, we 
describe the state-of-the-art models we
consider. We present the setup of our experiments in
Section~\ref{sec:setup} and their results in Section~\ref{sec:analysis}. Section~\ref{sec:relatedWork} discusses
relevant works, while
Section~\ref{sec:conclusions} concludes the paper along with directions for
future work.
%\vspace{-2pt}
%\vspace{-10pt}
\section{Preliminaries}
\label{sec:preliminaries}

\textbf{Representation Sources.}
We consider five sources of tweets for modeling the preferences of a Twitter
user, $u$:

\emph{(i)} The past \textit{retweets} of $u$, $R(u)$, which are the
tweets she has received from her followees and has reposted herself. Apparently,
their subjects have captured $u$'s attention so intensely that she decided
to % reproduce them in order to 
share them with her followers.

\emph{(ii)} All past \textit{tweets} of $u$ except her retweets,
$T(u)$. They enclose themes
$u$ is interested in chatting or in informing her followers. 

\emph{(iii)} All (re)tweets of $u$'s followees,
$E(u)=\bigcup_{u_{i}\in e(u)}(R(u_{i})\cup T(u_{i}))$,
where $e(u)=\{u_1,\dots,u_k\}$ is the set of her followees. $E(u)$
models a user as an \textit{information seeker}, who actively and explicitly follows
Twitter accounts that provide interesting information
\cite{DBLP:conf/icdm/KimS11,chen2010short,hannon2010recommending}.

\emph{(iv)} All (re)tweets of $u$'s followers,  
$F(u)=\bigcup_{u_{i}\in f(u)}(R(u_{i})\cup T(u_{i}))$, where
$f(u)=\{u_1,\dots,u_m\}$ stands for the set of her followers. 
%Ostensibly, $F(u)$ is inappropriate for modeling $u$'s preferences, as she
%exerts little control on her followers \cite{hannon2010recommending}.
Given that they have actively decided to follow $u$, due to the
interest they find in her posts, $F(u)$ models $u$ as an \textit{information producer} \cite{hannon2010recommending,ramage2010characterizing,DBLP:conf/webi/ZhaoT14}.

\emph{(v)} All (re)tweets of $u$'s reciprocal connections, 
$C(u)$=$E(u)$$\cap$$F(u)$. Unlike the unilateral following relationship in
Twitter, reciprocal connections may indicate users with very high affinity, thus
providing valuable information for user modeling.

Note that all these \textit{atomic} representation sources are complementary,
as they cover different aspects of the activity of a particular user and her network. For this reason, $T(u)$  is typically combined with $R(u)$ \cite{chen2012collaborative, DBLP:conf/wsdm/FengW13, DBLP:journals/tist/ShenWZHYSG13, lu2012twitter}, with only rare exceptions like \cite{DBLP:journals/tkde/JiangCWZY14}, which considers $T(u)$ in isolation. In this work, we consider not only $T(u) \cup R(u)$ (\textsf{TR} for short), but also
%We also consider 
the seven remaining
%possible 
pairwise combinations, which give
rise to the following \textit{composite} representation sources: 
$T(u) \cup E(u)$, $R(u) \cup E(u)$, $E(u) \cup F(u)$,
$T(u) \cup F(u)$, $R(u) \cup F(u)$, $T(u) \cup C(u)$, and $R(u) \cup C(u)$.
For simplicity, we denote them by \textsf{TE}, \textsf{RE}, \textsf{EF}, \textsf{TF}, \textsf{RF}, \textsf{TC}~and~\textsf{RC},~respectively.%in the following.

\textbf{Twitter Challenges.}
Tweets have some special characteristics that distinguish them from other
conventional domains and pose major challenges to representation models
\cite{chen2012collaborative,mehrotra2013improving,DBLP:journals/www/0001GP16}.

\emph{(C1)} \textit{Sparsity.} Tweets are short, comprising up to 280 characters for most languages, except for Chinese, Korean and Japanese, where the length limit is 140 characters.
  As a result, they lack sufficient content for user and document modeling.
 
\emph{(C2)} \textit{Noise.} The real-time nature of Twitter forces users to
tweet quickly, without taking into account the frequent grammar errors and misspellings; these
%quite frequent mistakes 
are corrected in subsequent tweets. 
% Inevitably, this yields high levels of noise.

\emph{(C3)} \textit{Multilingualism.} The global popularity of Twitter has
led to a high diversity in tweet languages. This renders inapplicable most 
language-specific pre-processing techniques, such as stemming and lemmatization. 
% as they rely on language characteristics in order to reduce words to their
% common root. 
Even tokenization becomes difficult: unlike the European ones, 
many Asian languages, such as Chinese, Japanese, and Korean, do not use spaces or other punctuation in order to distinguish consecutive words.

\emph{(C4)} \textit{Non-standard language.} Tweets offer
an everyday informal communication, which is unstructured, ungrammatical or
simply written in slang; words are abbreviated (e.g., ``gn'' instead of
``goodnight''), or contain emoticons, such as :), hashtags like \#edbt and
emphatic lengthening (e.g., ``yeeees'' instead of ``yes'').
 
We consider the above challenges when discussing the outcomes of our
experimental analysis in Section \ref{sec:analysis}.
%  in order to explain if and how every representation model tackles them.

\textbf{User Categories.} Twitter users are typically classified into three
categories \cite{armentano2011topology,java2007we}:
\emph{(i)} \textit{Information Producers} (\textsf{IP}) are those users who tweet and retweet more frequently than they receive updates from
their followees, (ii) \textit{Information Seekers} (\textsf{IS}) are those
users who are less active compared to their followees, and (iii)
\textit{Balanced Users} (\textsf{BU}) are those exhibiting a symmetry between
the received and the posted messages. 
%In our experiments, we combine all user
%categories with all representation models.

To quantify these categories, we use the ratio of
\textit{outgoing} to \textit{incoming tweets}. For a particular user $u$, the former involve the
tweets and retweets she posts, $R(u) \cup T(u)$, while the latter comprise the tweets
and retweets of her followees, $E(u)$. Dividing the outgoing with the incoming
tweets, we get the \textit{posting ratio}. Apparently, \textsf{BU} is
the set of users with a posting ratio close to 1, i.e., $|R(u) \cup
T(u)| \simeq |E(u)|$. To ensure significantly different behavior for the
other two categories, we define \textsf{IP} as the set of users with a posting
ratio higher than 2, thus indicating that they post at least twice as many
tweets as they receive. Symmetrically, we define \textsf{IS} as the set of users
with a posting ratio lower than 0.5, receiving at least twice as many tweets as
those they publish.
%themselves. 

\begin{figure}[t]\centering
\includegraphics[width=0.49\textwidth]{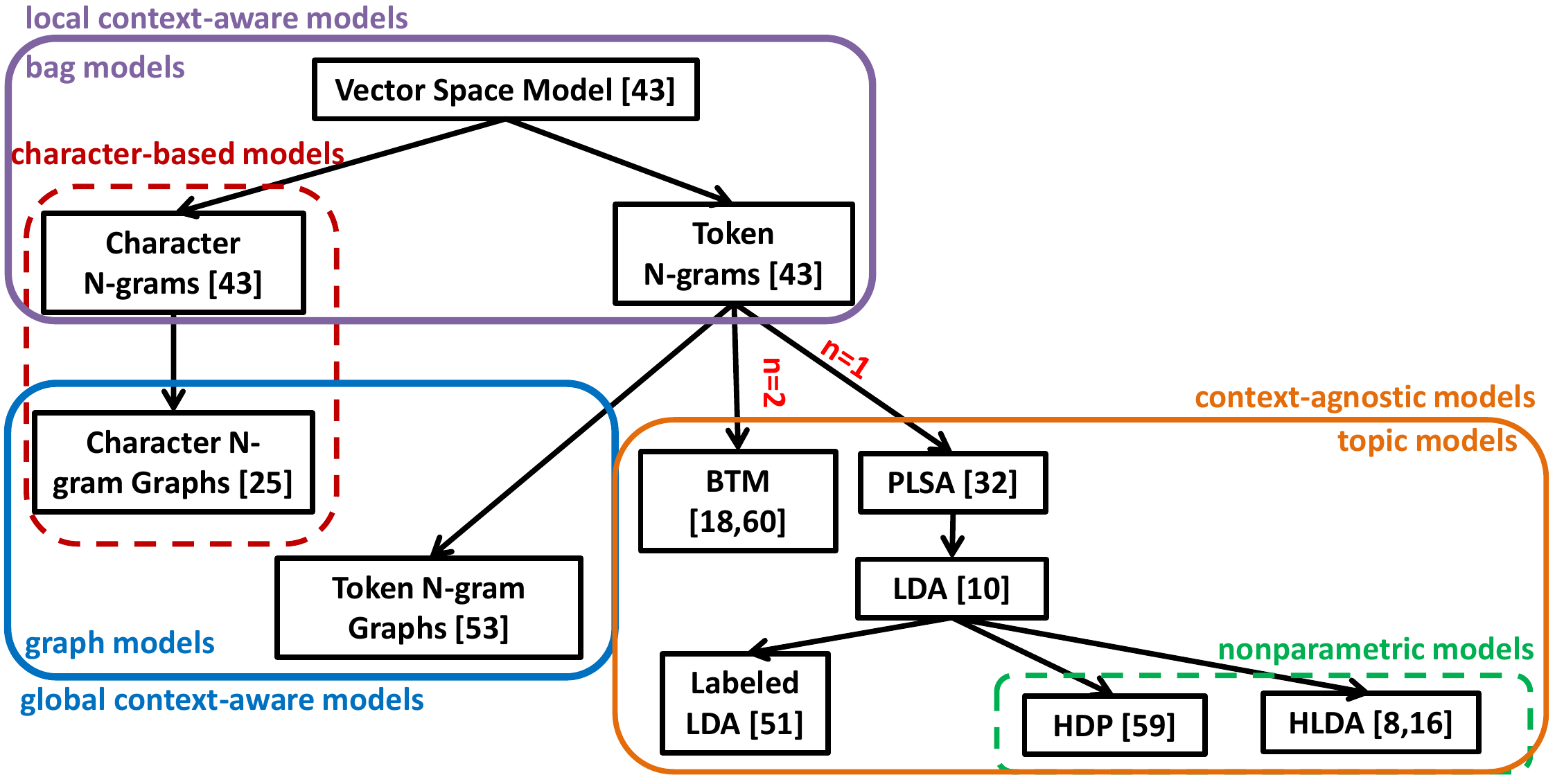}
\vspace{-18pt}
\caption{Taxonomy of representation models.}
\label{taxonomy-fig}
\vspace{-0.2in}
\end{figure}

\textbf{Problem Definition.} The task of Content-based PMR aims to distinguish a user's incoming messages, \textsf{R}$\cup$\textsf{T}, into
\emph{irrelevant} and \emph{relevant} ones. A common assumption in the literature \cite{chen2012collaborative,DBLP:journals/tist/ShenWZHYSG13,lu2012twitter, hong2010empirical}, which allows for large-scale evaluations, is that the relevant messages are those that are
retweeted by the user that receives them, an action that implicitly indicates
her interests -- intuitively, a user forwards a tweet to her followers after
carefully reading it and appreciating its content. 

In this context, Content-based PMR is usually addressed through a 
\textit{ranking-based recommendation algorithm}
\cite{DBLP:conf/ecir/ChenWHY14,chen2012collaborative,duan2010empirical}, 
which aims to rank relevant posts higher than the irrelevant ones. More
formally, let $D$ denote a set of documents, $U$ a set of users and $M$ the
representation space, which is common for both users and microblog posts (e.g., $M$ could be the real vector space). Given a user model $UM : U{\rightarrow}M$ and a document model $DM : D{\rightarrow}M$, we define this type of recommendation algorithms as follows:

%\vspace{-0.1in}
\noindent
\begin{definition}\textbf{Content-based Personalized Microblog Recommendation} learns a
ranking model $RM: M \times M \rightarrow \mathbb{R}$, which, given a user $u$
and a set of testing posts $D_{test}(u){=}\{d_1, \dots ,d_k\}{\subseteq}E(u)$, 
calculates the ranking scores $RM(UM(u), DM(d_i)), \forall i \in \{1 \dots k\}$,
and returns a list with %all documents in
$D_{test}(u)$ ranked in~decreasing~score.
\label{def:tweetRecommendationR}
\end{definition}
%\vspace{-0.1in}

We consider a user $u$ as the set of documents that stem from a particular representation source $s$, i.e., $s(u)$, and we build a different $UM_s(u)$ for each $s$. Given a set of users $U$ along with a representation source $s=\{s(u): u\in U\}$ and a set of labeled data $D_{tr}^s=\{D_{tr}(u): u \in U\}$, where $D_{tr}^s(u)=\{(d_i, l_i), d_i \in s(u), l_i \in L\}$ and $D_{tr}^s(u) \cap D_{test}(u) = \emptyset$, the recommendation algorithm is trained as follows for each individual source $s$: (i) for each $u \in U$, we learn $UM_s(u)$ and $DM(d_i)$ for each $d_i \in D_{tr}^s(u)$, and (ii) we train $RM$ on $\bigcup_{u \in U}\{(UM_s(u),DM(d_i)): i \in \{1, \dots
|D_{tr}^s(u)|\}$.

Note that 
%the overall performance of a recommendation approach mainly depends on three design choices: \emph{(i)} the evidence used by the user model, \emph{(ii)} the evidence used by the document model, and \emph{(iii)} the configuration of the recommendation algorithm. In this work, we focus on the first two, especially on their text-based features. 
as a recommendation algorithm, we employ the one commonly used in the literature for ranking-based PMR \cite{chen2010short,hannon2010recommending,kywe2012recommending,lu2012twitter}: in essence, $RM$ is a similarity function and $RM(UM(u), DM(d_i))$ quantifies the similarity of $d_i$'s document model with $u$'s~user~model. %As a result, the training set $D_{tr}$ of labeled incoming tweets is not needed, since there are no parameters to be learned.

%Note also that an alternative approach is to address (text-based) PMR through
%a binary classification algorithm \cite{DBLP:conf/aaai/BasuHC98}, which
%aims to predict whether a test document is relevant or not, associating it with
%a label from the set $L= \{
%relevant, irrelevant \}$. Yet, we do not consider this approach in our analysis
%in order to avoid the impact of classifier's fine-tuning on our experimental results.

%\vspace{-5pt}
%\vspace{-10pt}
%\input{taxonomy}
%\vspace{-10pt}
\section{Representation Models}
\label{sec:representationModels}

\subsection{Taxonomy} 
\label{sec:taxonomy}

We first present a taxonomy of the state-of-the-art representation models based on
their internal functionality, i.e., the way they handle the order of n-grams.
%on the structures that determine the information they enclose. 
Figure~\ref{taxonomy-fig} presents an
overview of their relations, with every edge $\mathbf{A\rightarrow B}$
indicating that model $B$ shares the same n-grams 
%of the same granularity
%same 
%core structures 
with $A$, but uses them in a
different way when representing users and documents. Based on these
relations, we identify three main categories of models:

\emph{(i)} \textbf{Context-agnostic models} disregard the order of
n-grams that appear in a document, when building its representation.
E.g., the token-based models of this category yield the same
representation for the phrases ``Bob sues Jim'' and ``Jim~sues~Bob''.
% Thus, they cannot
% % without consider into account the n-grams ordering inside the text, thus not 
% capture multiword expressions.

\emph{(ii)} \textbf{Local context-aware models} take into account 
the relative order of characters or tokens in the n-grams that lie at the core
of their document representations. 
% built document
% representations on word (or character) sequences, between words (or
% characters). However, 
Yet, they lose contextual information, as they ignore the ordering of n-grams
themselves. Continuing our example, the token-based models of this category are
able to distinguish the bigram ``Bob sues'' from ``sues Bob'', but cannot
capture the bigrams that precede or follow it.

\emph{(iii)} \textbf{Global context-aware models} incorporate into their
representations both the relative ordering of tokens or characters in
an n-gram and the overall ordering between n-grams in a document.
Continuing our example, models of this category distinguish ``Bob sues''
from ``sues Bob'' and know that the former is followed by ``sues Jim'', while
the latter is preceded by~``Jim~sues".

The first category comprises the \textit{topic models} \cite{blei2009topic}. These are
representation models that uncover the latent semantic structure of texts by
determining the topics they talk about. The topics and their proportions in a
document are considered as the hidden structure of a topic model, which can be
discovered by exclusively analyzing the observed data, i.e., the individual
tokens (words) in the original text. In general, they regard each document as 
a mixture of multiple topics, where each topic constitutes a set of co-occurring words. As a result, topic models are typically described in terms of probabilistic modeling, i.e., as generative processes that produce documents, words and users from distributions over the inferred topics \cite{blei2012probabilistic}.
%In this work, we exclusively consider \textit{probabilistic topic models} \cite{blei2012probabilistic}, which represent each document as a
%distribution over topics and, in turn, each topic as a distribution of words.

A subcategory of context-agnostic (topic) models pertains to \textit{nonparametric
models} \cite{hierarchicalTopicModels,teh2012hierarchical}, which adapt their representations to the structure of the training
data. They allow for an unbounded number of parameters that grows with the
size of the training data, whereas the parametric models are restricted
to a parameter space of fixed size. For example, the nonparametric topic models
assume that the number of topics is a-priori unknown, but can be inferred from
the documents themselves, while the parametric ones typically receive a fixed
number of topics as an input parameter before training.

The category of local context-aware models encompasses the bag models \cite{DBLP:books/daglib/0021593}, which
impose a strict order within n-grams: every n-gram is formed by a specific sequence of
characters or tokens and, thus, two n-grams with different sequences are
different, even if they involve the same characters or tokens; for example, the
bigrams ``ab'' and ``ba'' are treated as different. The only exception is the
token-based vector model with $n$=1, which essentially considers individual
words; its 
%Apparently, it involves a 
context-agnostic functionality
%, which 
actually lies at the core of most topic models.

Finally, the category of global context-aware models includes 
%are represented in our analysis
%by 
the n-gram graph models, which represent every document as a graph in a 
language-agnostic way \cite{giannakopoulos2008summarization}: every node corresponds to an~n-gram and edges connect pairs of n-grams co-located within a window of size $n$, with their weights indicating the co-occurrence frequency. These weighted edges allow graph models to capture global context, going beyond the local context of the bag models that use the same n-grams. Recent works suggest that the graph models outperform their bag counterparts in various tasks \cite{graphsTutorial}, which range from Information Retrieval \cite{DBLP:conf/cikm/RousseauV13} to Document Classification \cite{DBLP:journals/www/0001GP16}.
%\cite{giannakopoulos2008summarization}. They incorporate global contextual information, as they associate with
%edges all n-grams that co-occur in a window of size $n$ in the original text.
%Other instances of this category, not examined herein, are skip language models
%\cite{guthrie2006closer} and Hidden Markov Models (HMMs), which represent local
%phenomena (co-occurrence, neighborhood) and their statistics.

It should be stressed at this point that the bag and graph models share the
second subcategory of our taxonomy: the \textit{character-based models}. These
operate at a finer granularity than their token-based counterparts, thus being
more robust to noise \cite{DBLP:journals/www/0001GP16}. For example, consider the words
``tweet'' and ``twete'', where the second one is misspelled; they are
considered different in all types of token-based models, but for 
character bi-gram models, they match in three out of four bigrams. In this way,
the character-based models capture more precisely the actual similarity
between noisy documents.

In the following, 
%we refer to both token- and character-based n-gram graphs as
%\textit{graph models} for brevity. We also 
we collectively refer to local and
global context-aware models as \textit{context-based models}.

\begin{table}[t]\centering
\caption{The notation describing topic models.\label{tb:notation}}
\vspace{-8pt}
{\footnotesize
\begin{tabular}{ | l | l | }
\hline
\multicolumn{1}{|c}{\textbf{Symbol}} &
\multicolumn{1}{|c|}{\textbf{Meaning}} \\
\hline
\hline
$D$ & the corpus of the input documents \\
$|D|$ & the number of input documents \\
$d$ & an individual document in $D$ \\
$N_d$ & the number of words in $d$, i.e., the \textit{document length}\\
$U$ & the set of users\\
$u$ & an individual user in $U$\\
$N_{d,u}$ & the number of words in document $d$ of user $u$\\
$D_u$ & the documents posted by a user $u$ \\
$|D_u|$ & the number of documents posted by user $u$ \\
\hline
$V$ & the \textit{vocabulary} of $D$ (i.e., the set of words it includes) \\
$|V|$ & the number of distinct words in $D$ \\
$w$ & an individual word in $V$ \\
$w_{d,n}$ & the word at position $n$ in $d$ \\
\hline
$Z$ & the set of latent topics \\
$|Z|$ & the number of latent topics \\
$z$ & an individual topic in $Z$ \\
$z_{d,n}$ &the topic assigned to the word at position $n$ in document $d$ \\
\hline
$\theta_d$ & the multinomial distribution of document $d$ over $Z$, $\{P(z|d)\}_{z \in Z}$\\
$\theta_{d,z}$ & the probability that topic $z$ appears in document $d$, $P(z|d)$ \\
$\phi_z$ & the multinomial distribution of topic $z$ over $V$, $\{P(w|z)\}_{w \in V}$ \\
$\phi_{z,w}$ & the probability that word $w$ appears in topic $z$, $P(w|z)$ \\
$Dir(\alpha)$ & a symmetric Dirichlet distribution parameterized by $\alpha$ \\
\hline
\end{tabular}
}
%\vspace{-7pt}
\vspace{-17pt}
\end{table}

\subsection{State-of-the-art Models}

\begin{figure}[t!]\centering
    \includegraphics[width=0.49\textwidth]{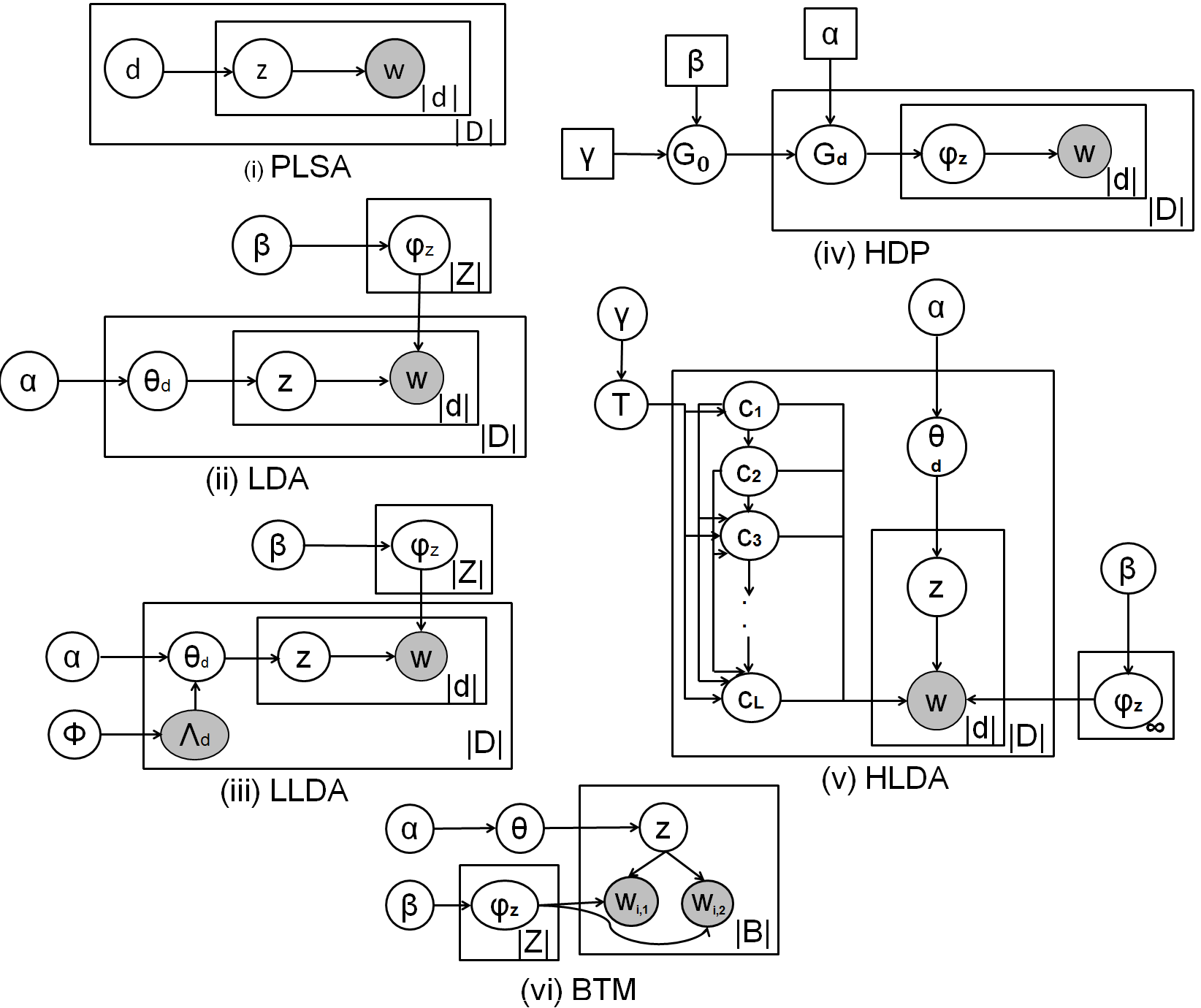}
    \vspace{-20pt}
    \caption{{\small Plate diagrams for: 
    (i) Probabilistic Latent Semantic Analysis (\textsf{PLSA}), 
    (ii) Latent Dirichlet Allocation (\textsf{LDA}),
    (iii) Labeled LDA (\textsf{LLDA}), 
    (iv) Hierarchical Dirichlet Process (\textsf{HDP}), 
    (v) Hierarchical LDA (\textsf{HLDA}), 
    and (vi) Biterm Topic Model (\textsf{BTM}).}}
    \vspace{-14pt}
\label{fig:plateDiagrams}
\end{figure}

We now elaborate on the main representation models that are widely used in the literature. 
%We begin with the bag models, the cornerstone of all representation models, and continue with their extension into graph and probabilistic topic models. 
To describe topic models, we use the common notation that is outlined in Table~\ref{tb:notation}, while their generative processes are illustrated in Figures \ref{fig:plateDiagrams}(i)-(vi) using \textit{plate diagrams}: shaded nodes correspond to the observed variables, the unshaded ones to the hidden variables, the arrows connect conditionally dependent variables, and finally, the plates denote repeated sampling for the enclosed variables for as many times as the number in the right bottom corner.

\textbf{Bag Models \cite{DBLP:books/daglib/0021593}}.
There are two types of n-grams, the character and the token
ones. These give rise to two types of bag models: the \textit{character n-grams
model} (\textbf{\textsf{CN}}) and the \textit{token n-grams model} (\textbf{\textsf{TN}}).
Collectively, they are called \textit{bag} or \textit{vector space models},
because they model a document $d_i$ as a vector with one dimension
for every distinct n-gram in a corpus of documents $D$: $DM(d_i)=(w_{i1},\dots,w_{im})$, 
where $m$ stands for the \textit{dimensionality} of $D$ (i.e., the number of distinct
n-grams in it), while $w_{ij}$ is the weight of the $j^{th}$ dimension that
quantifies the importance of the corresponding n-gram for $d_i$. 

The most common weighting schemes are:

\emph{(i)} \textit{Boolean Frequency} (\textbf{\textsf{BF}}) assigns binary weights that
indicate the absence or presence of the corresponding n-gram in $d_i$. 
More formally, $BF(t_j, d_i)$=1 if the n-gram $t_j$ of the $j^{th}$
dimension appears in document $d_i$, and 0 otherwise.

\emph{(ii)} \textit{Term Frequency} (\textbf{\textsf{TF}}) sets weights in proportion
  to the number of times the corresponding n-grams appear in document $d_i$. 
More formally, $TF(t_j, d_i)$=$f_j/N_{d_i}$, where $f_j$ stands for the
occurrence frequency of $t_j$ in $d_i$, while $N_{d_i}$ is the number of n-grams in $d_i$,
normalizing \textsf{TF} so as to mitigate the effect of different document
lengths on the weights.

\emph{(iii)}\textit{Term Frequency-Inverse Document Frequency}
(\textbf{\textsf{TF-IDF}}) discounts the \textsf{TF} weight for the most common
tokens in the entire corpus $D$, as they typically correspond to noise (i.e.,
stop words). Formally, $TF$-$IDF(t_j, d_i)=TF(t_j, d_i)\cdot IDF(t_j)$,
where $IDF(t_j)$ is the inverse document frequency of the n-gram $t_j$, i.e.,
$IDF(t_j)=\log{|D|/(|\{ d_k \in D : t_j \in d_k \}|+1)}$. In this way, high
weights are given to n-grams with high frequency in $d_i$, but low frequency in $D$.

To construct the bag model for a specific user $u$, we aggregate the vectors
corresponding to the documents that capture $u$'s interests. The end result is
a weighted vector $(a(w_{1}),....,a(w_{m}))$, where $a(w_{j})$ is the \textit{aggregation function} that calculates the importance
of the~$j^{th}$~dimension for $u$. 

The main aggregation functions are:

\emph{(i)} the sum of weights, i.e., $a(w_{j})=\sum_{d_i \in D} w_{ij}$.

\emph{(ii)} the centroid of unit (normalized) document vectors, i.e., 
$a(w_{j})=\frac{1}{|D|}\cdot\sum_{d_i \in D} \frac{w_{ij}}{||DM(d_i)||}$, where $||DM(d_i)||$ is the magnitude of $DM(d_i)$.

\emph{(iii)} the Rocchio algorithm,
%  \footnote{The Rocchio algorithm was initially proposed for relevance feedback in the vector-space model, but was later adapted to text classification \cite{sebastiani2002machine}. In this context, it represents each class $c_i$ of a set of classes $C$ as a prototype vector, which is constructed by aggregating the positive and negative examples for that specific class. To classify an unseen document, it computes the similarity between all class vectors and the document vector and selects the most similar class. Likewise, a user model can be built by combining positive and negative training data and then, the testing documents are compared with that model.} 
i.e.,\\
{\footnotesize
 $a(w_{j})=\alpha/|D^p| \cdot \sum_{d_i\in
 D^p} w_{ij}/||DM(d_i)||- \beta/|D^n| \cdot \sum_{d_i\in
 D^n} w_{ij}/||DM(d_i)||$}, where $D^p$ and $D^n$ are the 
 sets of positive (relevant) and negative (irrelevant) documents in $D$,
 respectively, while $\alpha, \beta \in [0,1]$ control the relative importance of
 positive and negative examples, respectively, such that $\alpha + \beta = 1.0$ \cite{DBLP:books/daglib/0021593}. 
 
To compare two bag models, $DM(d_i)$ and $DM(d_j)$, one of the following similarity
measures is typically used:

%{\color{red}\emph{(i)} \textit{Dot Product Similarity} (\textbf{\textsf{DPS}}) is the sum of the products
%of the corresponding weights of the two vectors. Formally:\\
%$DPS(DM(d_i),DM(d_j))=\sum_{k=1}^{m}{w_{ik}w_{jk}}$.}

\emph{(i)} \textit{Cosine Similarity} (\textbf{\textsf{CS}}) measures the cosine of the angle
 of the weighted vectors. Formally, it is equal to their dot product
 similarity, normalized by the product of their magnitudes:\\
$CS(DM(d_i),DM(d_j))=\sum_{k=1}^{m}{w_{ik}w_{jk}}/||DM(d_i)||/||DM(d_j)||$.
 
\emph{(ii)} \textit{Jaccard Similarity} (\textbf{\textsf{JS}}) treats the
document vectors as sets, with weights higher than (equal to) 0 indicating the
presence (absence) of the corresponding n-gram. On this basis, it defines as
similarity the ratio between the sizes of set intersection and union:
$JS(DM(d_i),DM(d_j)){=}|DM(d_i){\cap}DM(d_j)|/|DM(d_i){\cup}DM(d_j)|$.

\emph{(iii)} \textit{Generalized Jaccard Similarity} (\textbf{\textsf{GJS}}) extends
\textsf{JS} so that it takes into account the weights associated with every
n-gram:\\
{\small$GJS(DM(d_i),DM(d_j)){=}\sum_{k=1}^{m}min(w_{ik},w_{jk})/\sum_{k=1}^{m}max(w_{ik},w_{jk})$}. \\
Note that for \textsf{BF} weights, \textsf{GJS} is identical with \textsf{JS}. 

\textbf{Graph Models \cite{giannakopoulos2008summarization,DBLP:conf/cikm/RousseauV13}.}
%Recent works suggest that graph models outperform the bag ones in various tasks \cite{graphsTutorial}, from Information Retrieval \cite{DBLP:conf/cikm/RousseauV13} to Document Classification \cite{DBLP:journals/www/0001GP16}. 
%Therefore, it is worth applying them to \textsf{PMR}, too.
There are two graph models, one for each type of n-grams, i.e.,
\textit{token n-gram graphs} (\textbf{\textsf{TNG}}) \cite{DBLP:conf/cikm/RousseauV13} and
\textit{character n-gram graphs} (\textbf{\textsf{CNG}}) \cite{giannakopoulos2008summarization}. Both models represent each document $d$ as an undirected graph
$G_{d}$ that contains one vertex for each n-gram in $d$.
An edge connects every pair of vertices/n-grams that
co-occur within a window of size $n$ in $d$. Every 
edge is weighted according to the co-occurrence frequency of the
corresponding n-grams. Thus, the graphs
incorporate \textit{contextual information} in the form of n-grams'
closeness. 

To construct the model for a user $u$, we merge the graphs of the 
documents representing $u$'s interests using the update
operator, which is described in~\cite{giannakopoulos2010content}.
To compare graph models, we can use the following graph similarity
measures~\cite{giannakopoulos2008summarization}:

\emph{(i)} \textit{Containment Similarity} (\textbf{\textsf{CoS}}) estimates the number
of edges shared by two graph models, $G_i$ and $G_j$, regardless of the
corresponding weights (i.e., it merely estimates the portion of common n-grams in the original
texts). 
%, which corresponds to \textsf{CS} with \textsf{BF} weights for bag models. 
Formally: $CoS(G_i,G_j) = \sum_{e\in G_i}{\mu(e,G_j)}/min(|G_i|,|G_j|)$,
where $|G|$ is the size of graph G, and $\mu(e,G)=1$ if $e \in G$, or 0
otherwise.

\emph{(ii)}  \textit{Value Similarity} (\textbf{\textsf{VS}}) 
extends \textsf{CoS} by considering the weights of common edges. Formally, using $w_e^k$ for the weight of edge $e$ in $G_k$:
{\small
$VS(G_i,G_j)=\sum_{e\in (G_i\cap G_j)}{\frac{min(w_e^i,w_e^j)}{max(w_e^i,w_e^j)\cdot max(|G_i|,|G_j|)}}$}.
%Thus, \textsf{VS} corresponds to \textsf{CS} with \textsf{TF} weights for bag models.

\emph{(iii)} \textit{Normalized Value Similarity} (\textbf{\textsf{NS}}) extends
\textsf{VS} by mitigating the impact of imbalanced graphs, i.e., the cases where
the comparison between a large graph with a much smaller one yields
similarities close to 0. Formally:\\
{\small
$NS(G_i,G_j){=}\sum_{e\in
(G_i\cap G_j)}{min(w_e^i,w_e^j)/max(w_e^i,w_e^j)}/min(|G_i|,|G_j|)$}.

\textbf{Probabilistic Latent Semantic Analysis (\textsf{PLSA})
\cite{hofmann1999probabilistic}.}
This model assigns a topic $z$ to every observed word $w$ in a document $d$.
Thus, every document is modeled as a distribution over multiple topics, assuming
that the observed variables $w$ and $d$ are conditionally independent given the
unobserved topic $z$: $P(w|d,z) = P(w|z)$. For an observed pair $(w,d)$, the
joint probability distribution~is:
$P(w,d) = P(d){\cdot}{\sum_z}{P(w,z|d)} = P(d){\cdot}{\sum_z}{P(z|d){\cdot}P(w|z)} = P(d){\cdot}{\sum_z}{\theta_{d,z}{\cdot}\phi_{z,w}}$,
where $\theta_{d,z}$ stands for the probability that a topic $z$
appears in document $d$, while $\phi_{z,w}$ denotes the probability that a
word $w$ appears in topic $z$ (see Table \ref{tb:notation}).

Figure~\ref{fig:plateDiagrams}(i) depicts the generative process of \textsf{PLSA}:
\emph{(1)} Select a document $d$ with probability $P(d)$.
\emph{(2)} For each word position $n \in \{1, \dots ,N_d\}$:
\emph{(a)} Select a topic $z_{d,n}$ from distribution $\theta_d$.
\emph{(b)} Select the word $w_{d,n}$ from distribution $\phi_{z_{d,n}}$. 
Note that $P(d)$ is the frequency of $d$ in the corpus, thus being uniform in
practice.

In total, \textsf{PLSA} should estimate $|D|\cdot|Z|+|Z|\cdot|V|$ parameters:
 $\theta_d=\{P(z|d)\}_{z \in Z}$ for each $d\in D$ and $\phi_z=\{P(w|z)\}_{w \in V}$ 
for each $z\in Z$. Consequently, the number of parameters grows linearly with the
number of documents, leading to overfitting~\cite{blei2003latent}.

\textbf{Latent Dirichlet Allocation (\textsf{LDA}) \cite{blei2003latent}.}
Unlike \textsf{PLSA}, which regards each document $d$ as a 
list of probabilities $\theta_d$, \textsf{LDA} assigns a random variable
of $|Z|$ parameters with a Dirichlet prior to distribution $\theta_d$. 
In a latter variant, a $|V|$-parameter variable with a Dirichlet prior was also
assigned to $\phi_z$ \cite{griffiths2004finding}. 
The number of topics $|Z|$ is given as a parameter to the model and raises the following issue: the smaller the number of topics is, the broader and more
inaccurate is their content, failing to capture the diverse themes discussed in the
corpus; in contrast, for large values of $|Z|$, the model is likely to overfit, learning spurious word co-occurrence patterns~\cite{steyvers2007probabilistic}.

Figure \ref{fig:plateDiagrams}(ii) shows the generative process of \textsf{LDA}:
\emph{(1)} For each topic $z \in Z$, draw a distribution $\phi_z$ from $Dir(\beta)$. 
\emph{(2)} For each document $d \in D$: \emph{(a)} Select a distribution $\theta_d$ from $Dir(\alpha)$. \emph{(b)} For each word position $n \in \{1, \dots ,N_d\}$: \emph{(i)} Draw a topic $z_{d,n}$ from distribution $\theta_d$. \emph{(ii)} Draw the word $w_{d,n}$ from distribution $\phi_{z_{d,n}}$.

Note that the hyperparameters $\alpha$ and $\beta$ of
the Dirichlet priors on $\theta$ and $\phi$, respectively, distinguish
\textsf{LDA} from \textsf{PLSA}. The former denotes the frequency with which a
topic is sampled for a document, while the latter shows the frequency of a word
in a topic, before actually observing any words in $D$.

\textbf{Labeled LDA (\textsf{LLDA}) \cite{ramage2009labeled}.}
This is a supervised variant of \textsf{LDA} that characterizes a corpus $D$
with a set of observed labels $\Lambda$. Each document $d$ is modeled as a multinomial
distribution of labels from $\Lambda_d \subseteq \Lambda$. Subsequently,
each word $w \in d$ is picked from a distribution $\phi_z$ of some label
$z$ contained in $\Lambda_d$. Besides the observed labels, \textsf{LLDA} can also use $|Z|$ latent topics for
all documents, assigning the labels "Topic $1$",\ldots, "Topic $|Z|$" to
each document $d \in D$ \cite{ramage2010characterizing}.

Figure \ref{fig:plateDiagrams}(iii) presents the generative process of \textsf{LLDA}: \emph{(1)} For each topic $z \in Z$, draw a distribution $\phi_z$ from
  $Dir(\beta)$.   \emph{(2)}  For each document $d \in D$:
  \emph{(a)}  Construct distribution $\Lambda_d$ by selecting each topic $z \in Z$ 
    as a label based on a Bernoulli distribution with parameter $\Phi_{z}$.
  \emph{(b)}  Select a multinomial distribution $\theta_d$ over 
 $\Lambda_d$ from $Dir(\alpha)$.
\emph{(c)}  For each word position $n \in \{1, \dots ,N_d\}$:    
 	 \emph{(i)} Draw a label $z_{d,n}$ from distribution $\theta_d$.
 	   \emph{(ii)} Draw the word $w_{d,n}$ from distribution $\phi_{z_{d,n}}$. 
Note that the prior probability of adding a topic $z$ to $\Lambda_d(\Phi_z)$
is practically superfluous, as $\Lambda_d$ is observed for each document $d$.

\textbf{Hierarchical Dirichlet Process (\textsf{HDP})
\cite{teh2012hierarchical}.}
This Bayesian nonparametric model is crafted for
clustering the observations of a group into mixing components. 
In PMR, each document corresponds to a group, the words of
the document constitute the observations within the group, and the topics
comprise the mixing components in the form of distributions over words.

Two are the main properties of \textsf{HDP}: \emph{(i)} The number of mixing components
is countably infinite and unknown beforehand. This is achieved by assigning a
random variable $G_j$ to the $j^{th}$ group distributed according to
$DP(\alpha,G_0)$, where $DP$ stands for a Dirichlet Process, which is a
probability distribution over distributions (i.e., samples from a DP are
probability distributions themselves). $G_0$ is the base probability
distribution, playing the role of the mean around which distributions are
sampled by $DP$, while $\alpha$ is called concentration parameter and can be
thought as an inverse variance. $G_0$ also follows a
$DP(\gamma,H)$.
\emph{(ii)} The groups share the same components. This is achieved by linking
the $DPs$ of all groups, under the same $G_0$.
More formally, \textsf{HDP} is defined as:
$G_0|\gamma,H\sim DP(\gamma,H)$ and $G_j|\alpha,G_0\sim DP(\alpha,G_0) \mbox{ }
\forall j$.

Figure \ref{fig:plateDiagrams}(iv) shows the generative process for 1 hierarchical level: 
\emph{(1)} Draw $G_0$ from  $DP(\gamma,H)$, where $H$ is a
$Dir(\beta)$. $G_0$ provides an unbounded number of $\phi_z$ distributions,
i.e., topics that can be assigned to any document $d \in D$ \cite{teh2012hierarchical}.
\emph{(2)} For each document $d \in D$:
\emph{(a)} Associate a subset of distributions $\phi_z$ with $d$, by
  	drawing $G_d$ from $DP(\alpha,G_0)$.
\emph{(b)} For each word position $n \in \{1, \dots ,N_d\}$: 
  		\emph{(i)} Pick a distribution $\phi_{z_{d,n}}$ from $G_d$.
  		\emph{(ii)} Draw the word $w_{d,n}$ from $\phi_{z_{d,n}}$.

Note that it is straightforward to add more hierarchical levels to \textsf{HDP},
by exploiting its recursive nature. For example, in multiple corpora, where
documents are grouped into broader categories,
topics shared between categories are revealed and can be compared with topics
shared between individual documents~\cite{teh2012hierarchical}.

\textbf{Hierarchical LDA (\textsf{HLDA}) \cite{griffiths2004hierarchical,hierarchicalTopicModels}.}
This model extends \textsf{LDA} by organizing the topics in $Z$ into a
hierarchical tree such that every tree node represents a single topic.
The broader a topic is, the higher is its level in the tree, with the most
specific topics assigned to the leaves. Although the tree levels are fixed, the branching factor is inferred
from the data, leading to a nonparametric functionality.
Each document is modeled as a path from the root to a leaf and its words are
generated by the topics across this path. Hence, every document representation
is derived from the topics of a single path, rather than all topics in $Z$
(as in \textsf{LDA}).

Specifically, \textsf{HLDA} is based on the \textit{Chinese Restaurant
Process} (CRP), which is a distribution over partitions of integers. CRP assumes the existence of a Chinese restaurant with an infinite number
of tables. The first customer selects the first table, while the $n^{th}$ customer selects a table based on the
following probabilities \cite{griffiths2004hierarchical}:
$\mathbf{P_1}$=$P($occupied $table_i$$|$previous
customers$)$=$n_i$$/$$(n$-1+$\gamma)$, \\
$\mathbf{P_2}$=$P($next
unoccupied $table_i$$|$previous customers$)$=$\gamma$$/$$(n$-1+$\gamma)$,\\ 
%\noindent
where $\gamma$
is a parameter controlling the possibility for a new customer to sit to an
occupied or an empty table, 
%or to an empty one, 
and $n_i$ are the customers already seated
on table $i$. The placement of $M$ customers produces a partition of $M$ integers.

In fact, \textsf{HLDA} relies on the \textit{nested Chinese Restaurant
Process} (nCRP), which extends CRP by building a infinite hierarchy of Chinese
restaurants. It assumes that there exists an infinite number of Chinese
restaurants, each with an infinite number of tables. One of the restaurants is
the root and every restaurant's table has a label, pointing to another
restaurant -- a restaurant cannot be referred by more than one label in the
hierarchy. To illustrate the functionality of this process, assume a customer that
visits the restaurants for $L$ days. Starting from the root, she forms a path of
$L$ restaurants, one per day, following the labels of the tables she has chosen
to sit based on $P_1$ and $P_2$. The repetition of this process for $M$
customers yields an $L$-level hierarchy of~restaurants.

Figure \ref{fig:plateDiagrams}(v) presents the generative process of
\textsf{HLDA}, with $T$ denoting an infinite tree drawn from
nCRP$(\gamma)$ (i.e., the infinite set of possible $L$-level paths):
\emph{(1)} For each restaurant $z$ in $T$, draw a distribution $\phi_z$ from
$Dir(\beta)$.
\emph{(2)} For each document $d \in D$:
  \emph{(a)} Draw an $L$-level path from $T$ as follows: Let $c_{d,1}$ be the root restaurant. For each level $l \in \{2, \dots, L\}$ pick a table from restaurant
$c_{d,l-1}$ using $P_1$ and $P_2$ and set $c_{d,l-1}$ to refer to the restaurant
$c_{d,l}$, which is indicated by that table.
    \emph{(b)} Select a distribution $\theta_d$ over $\{1, \dots ,L\}$ from
    $Dir(\alpha)$.
    \emph{(c)} For each word position $n \in \{1, \dots ,N_d\}$:
      \emph{(i)} Draw a level $l_{d,n} \in \{1, \dots, L\}$ from $\theta_d$.
      \emph{(ii)} Draw the word $w_{d,n}$ from  distribution $\phi_{z_{d,n}}$, 
      where $z_{d,n}$ is the topic corresponding to the~restaurant~$c_{d,l_{d,n}}$.
      
\textbf{Biterm Topic Model (\textsf{BTM}) \cite{cheng2014btm,yan2013biterm}.}
At the core of this model lies the notion of \textit{biterm}, which is an unordered pair of words that are located in close distance within a given text. In short documents, close distance means that both words belong to the same document, whereas in longer texts, it means that they co-occur within a window of tokens that is given as a parameter to \textsf{BTM}. Based on this notion, \textsf{BTM} addresses the sparsity in short texts like tweets (Challenge C1) in two ways: \emph{(i)} It models the word co-occurrence for topic learning \textit{explicitly} by considering biterms (rather than \textit{implicitly} as in \textsf{LDA}, where word co-occurrence is captured by drawing a document's words from topics of the same topic distribution $\theta_d$). \emph{(ii)} It considers the entire corpus $D$ as a set of biterms $B$, extracting word patterns from the entire training set (rather than an individual document). Overall, \textsf{BTM} assumes that the corpus consists of a mixture of topics and directly models the biterm generation from these topics. 

Figure \ref{fig:plateDiagrams}(vi) shows the generative process of  \textsf{BTM}: \emph{(1)} For each topic $z \in Z$, draw a distribution $\phi_z$ from
  $Dir(\beta)$.
  \emph{(2)} For the entire corpus $D$, select
a multinomial distribution $\theta$ over $Z$ from
$Dir(\alpha)$. 
  \emph{(3)} For each biterm position $n$ in the entire corpus $\{1, \dots ,|B|\}$:
  \emph{(a)} Draw a topic $z_n$ from $\theta$.
    \emph{(b)} Draw two words $w_{n,1}$, $w_{n,2}$ from~$\phi_{z_n}$. 

Note that
\textsf{BTM} does not contain a generation process for documents. The distribution
$\theta_d$ for an individual document $d \in D$ is inferred from the formula
$P(z/d)=\sum_{b\in d}{P(z/b)\cdot P(b/d)}$, which presumes that the
document-level topic proportions can be derived from the document-level generated biterms
\cite{cheng2014btm,yan2013biterm}.

\textbf{Other models.} There is a plethora of topic models in the literature. Most of them lie out of the scope of our experimental analysis, because they encapsulate external or non-textual information. For example, 
\cite{DBLP:conf/cikm/SangLX15} and \cite{DBLP:conf/wise/ChenZZC13} incorporate  temporal information from users' activity, while \cite{DBLP:conf/kdd/El-AriniPHGA12} proposes a representation model called \textit{badges}, which combines the posting activity of Twitter users with self-reported profile information. Other topic models are incompatible with the ranking-based recommendation algorithm for Content-based PMR. For example, Twitter-LDA \cite{zhao2011comparing} and Dirichlet Multinomial Mixture Model 
\cite{nigam2000text} assign a single topic to every tweet, thus yielding too many ties when ranking document models in decreasing similarity score - all tweets with the same inferred topic 
are equally similar with 
%have the same similarity with 
the user model. 

\textbf{Using Topic Models.} 
To build a topic model for a particular
user $u$, we average the distributions corresponding to the documents that
capture her preferences. To compare user models with
document models, we use the cosine similarity. 
To address the four Twitter challenges, which hamper the
functionality of topic models due to the scarcity of word co-occurrence
patterns, we apply two different pooling schemes to
the training data: \emph{(i)} the aggregation on users, called \textit{User
Pooling} (\textsf{UP}), where all tweets posted by the same user are considered as a
single document, and \emph{(ii)} the aggregation on hashtags, called \textit{Hashtag
Pooling} (\textsf{HP}), where all tweets annotated with the same hashtag form a
single document (the tweets without any hashtag are treated as individual
documents). We also consider \textit{No Pooling}
(\textsf{NP}), where each tweet is considered as an individual document.
Finally, we estimate the parameters of all topic models using \textit{Gibbs Sampling} \cite{geman1984stochastic}, except for \textsf{PLSA}, which uses \textit{Expectation Maximization} \cite{dempster1977maximum}.
%\vspace{-7pt}
\section{Experimental Setup}
\label{sec:setup}

\begin{table}[t]
%\centering
\caption{Statistics for each user group in our
dataset.}
\vspace{-8pt}
{\small
\begin{tabular}{ | l || r r r r |}
   \cline{2-5}
   \multicolumn{1}{c|}{} & \multicolumn{1}{c}{\textbf{\textsf{IS}}} &
   \multicolumn{1}{c}{\textbf{\textsf{BU}}} &
   \multicolumn{1}{c}{\textbf{\textsf{IP}}} & 
   \multicolumn{1}{c|}{\textbf{\textsf{All Users}}} \\
   \hline
   \hline
   Users & 20 & 20 & 9 & 60\\
   \hline
   \hline
   Outgoing tweets (\textsf{TR}) & 47,659	& 48,836 & 42,566 & 192,328
   \\
   \hline
   Minimum per user & 1,100 & 766 & 1,602 & 766
   \\
   Mean per user & 2,383 & 2,442 & 4,730 & 3,205
   \\
   Maximum per user & 6,406	& 8,025 & 17,761 & 17,761
   \\
   \hline
   \hline
    Retweets (\textsf{R}) & 27,344 & 32,951 & 38,013 & 140,649
    \\
   \hline
   Minimum per user & 840 & 445 & 1,198 & 445
   \\
   Mean per user & 1,367 & 1,648 & 4,224 & 2,344
   \\
   Maximum per user & 2,486 & 6,814 & 17,761 & 17,761
   \\
   \hline
   \hline
   Incoming tweets (\textsf{E}) & 390,638 & 49,566 & 10,285 & 484,698
   \\
   \hline
   Minimum per user & 8,936 & 696 & 525	& 525
   \\
   Mean per user & 19,532 & 2,478 & 1,143 & 8,078
   \\
   Maximum per user & 53,003 & 7,726 & 1,985 & 53,003
   \\
   \hline
   \hline
   Follower's tweets (\textsf{F}) & 665,778 & 166,233 & 50,330 & 1,391,579
   \\
   \hline
   Minimum per user & 1,074 & 110 & 348 & 110
   \\
   Mean per user & 33,289 & 8,312 & 5,592 & 23,193
   \\
   Maximum per user & 144,398 & 52,318 & 33,639 & 447,639
   \\
   \hline
   %\hline
   %Testing tweets & 2,520 & 2,261 & 170 & 7,041
   %\\
   %\hline
   %Minimum per user & 20 & 20 & 10 & 10
   %\\
   %Mean per user & 126 & 113 & 19 & 117
   %\\
   %Maximum per user & 410 & 860 & 50 & 1,075
   %\\
   %\hline
\end{tabular}
}
%\vspace{-7pt}
\label{tb:dataset}
\vspace{-10pt}
\end{table}

All methods were implemented in Java 8 and performed in a server with Intel Xeon
E5-4603@2.20 GHz (32 cores) and 120GB RAM, running Ubuntu 14.04. 

\begin{table*}[t]\centering
% {\fontsize{7.7pt}{7pt}\selectfont
% {\renewcommand{\arraystretch}{1.4}
\caption{The 10 most frequent languages in our dataset, which
collectively cover 1,879,470 tweets (91\% of all tweets).}
\vspace{-8pt}
{\small
\begin{tabular}{ | l | r | r | r | r | r |  r | r | r | r | r |}
\cline{2-11}
\multicolumn{1}{c|}{} &
\multicolumn{1}{c|}{English} & 
\multicolumn{1}{c|}{Japanese} & 
\multicolumn{1}{c|}{Chinese} & 
\multicolumn{1}{c|}{Portuguese} &
\multicolumn{1}{c|}{Thai} & 
\multicolumn{1}{c|}{French}	& 
\multicolumn{1}{c|}{Korean}	& 
\multicolumn{1}{c|}{German} & 
\multicolumn{1}{c|}{Indonesian} &
\multicolumn{1}{c|}{Spanish} \\
\hline
\hline
Total Tweets &
1,710,919 &
71,242 &
35,356 &
14,416 &
13,964 &
12,895& 
10,220& 
5,038& 
4,339& 
1,081\\
Relative Frequency &
82.71\% &	
3.44\% &
1.71\% &	
0.70\% &	
0.68\% &
0.62\% & 
0.49\% & 
0.24\% & 
0.21\% & 
0.05\% \\
\hline
\end{tabular}
}
%\vspace{-7pt}
\label{tb:languages}
\vspace{-10pt}
\end{table*}

Our dataset was derived from a large Twitter corpus 
%with
%more than 476 million tweets that have been posted by 20 million users from June
%1, 2009 to December 31, 2009 \cite{DBLP:conf/wsdm/YangL11}. Although slightly dated, this dataset 
that captures almost 30\% of all public messages published on Twitter worldwide between June
1, 2009 and December 31, 2009 \cite{DBLP:conf/wsdm/YangL11}.
%during that particular time frame. 
Although slightly dated,
%Besides, 
recent studies have verified that core aspects of the users activity in Twitter remain unchanged over the years (e.g., retweet patterns for individual messages and users) \cite{DBLP:conf/edbt/GrossettiCMT18}.
Most importantly, this dataset can be combined with a publicly available snapshot of the entire social graph of Twitter as of August,
2009 (\url{https://an.kaist.ac.kr/traces/WWW2010.html}). Given that every record includes the raw tweet along with the corresponding usernames and timestamps, we can simulate the tweet feed of every user during the second half of 2009 with very high accuracy. To ensure that there is a critical mass of tweets for all representation sources and a representative
set of testing documents for every user,  we removed from our dataset those users that had less than three
followers and less than three followees. We also discarded all users with less
than 400 retweets.

From the remaining users, we populated each of the three user types we defined
in Section~\ref{sec:preliminaries} with 20 users. For \textsf{IS}, we selected
the 20 users with the lowest posting ratios, for \textsf{BU} the 20 users with
the closest to 1 ratios, and for \textsf{IP}, the 20 users with the highest
ratios. The difference between the maximum ratio for \textsf{IS} (0.13) and 
the minimum one for \textsf{BU} (0.76) is large enough to guarantee
significantly different behavior. However, the maximum ratio among \textsf{BU} users is 1.16,
whereas the minimum one among \textsf{IP} users is 1.20, due to the scarcity
of information providers in our dataset. This means that the two
categories are too close, a situation that could introduce noise to the experimental results. To ensure
distinctive behavior, we placed in the \textsf{IP} group the nine
users that have a posting ratio higher than 2. The remaining 11 users with the
highest ratios are included in the group \textsf{All Users}, which additionally
unites \textsf{IS}, \textsf{BU} and \textsf{IP} to build a large dataset with 
2.07 million tweets and 60 users, in total.
% we also consider an additional user group, called \textit{Pure
% Information Producers} (\textsf{PIP}), that includes only those users with a posting ratio
% higher than 2.0; 9 users meet this requirement in our dataset. 
The technical characteristics of the four resulting user groups appear in Table~\ref{tb:dataset}. 
% In total, we employ more than 2 million tweets in our
% experiments.

Each user has a different train set, which is determined by the representation
source; for example, the train set for \textsf{T} contains all tweets of a user,
except for her retweets. In contrast, the test set of every user is
independent of the representation source and contains her incoming tweets; the
retweeted ones constitute the \textit{positive examples} and the rest are the
\textit{negative examples} \cite{chen2012collaborative,DBLP:journals/tist/ShenWZHYSG13,lu2012twitter, hong2010empirical}. However, the former are sparse and
imbalanced across time in our dataset. Following \cite{chen2012collaborative},
we retain a reasonable proportion between the two classes for each user by 
placing the 20\% most recent of her retweets in the test set. The
earliest tweet in this sample splits each user's timeline in two phases: the
\textit{training} and the \textit{testing phase}.
Based on this division, we sampled the negative data as follows: for each
positive tweet in the test set, we randomly added four negative ones from the
testing phase \cite{chen2012collaborative}.
Accordingly, the train set of every representation source is restricted to all the 
tweets that fall in the training phase.

Based on this setup, we build the user models as follows:
for the bag and graph models, we learn a separate model $UM_s(u)$ for every
combination of a user $u$ and a representation source $s$, using the
corresponding train set. For the topic models, we first learn a single model
$M(s)$ for each representation source, using the train set of all 60 users.
Then, we use $M(s)$ to infer the distributions over topics for the training
tweets of $u$ that stem from $s$. Finally, the user model $UM_s(u)$ is constructed by computing the centroid of
$u$'s training vectors/tweets \cite{ramage2010characterizing}.

Note that for all models, we converted the raw text of all training and testing tweets into lowercase. For all token-based models, we 
%token-based models, all training and testing tweets were
%pre-processed all tweets in the following way: 
%we converted the raw text into lowercase, 
%we 
tokenized  all tweets on white spaces and punctuation, we squeezed repeated letters 
and we kept together URLs, hashtags, mentions and emoticons. 
We also removed the 100 most frequent tokens across all training tweets, as they
practically correspond to stop words. We did not apply any language-specific
pre-processing technique, as our dataset is multilingual.
%(see below). 

%\textbf{Multilinguality.}
In more detail, 
%Our experimental methodology is language-agnostic in the sense that it does
%not require a-priori knowledge of the document languages. 
% Language detection lies out of the scope of this work, as it is a major
% challenge itself.
%Yet, it is worth examining the linguistic diversity of our dataset. 
Table \ref{tb:languages} presents the 10 most frequent languages in our dataset along
with the number of tweets that correspond to them. To identify them, we first
cleaned all tweets from hashtags, mentions, URLs and emoticons in order to
reduce the noise of non-English tweets. Then, we aggregated the tweets per
user (\textsf{UP}) to facilitate language detection. Finally, we automatically
detected the prevalent language in every pseudo-document (i.e., user)
\cite{langugageDetector} and
assigned all relevant tweets to it. As expected, English is the dominant
language, but our corpus is highly multilingual, with 3 Asian languages ranked
within the top 5 positions. This multilingual content (Challenge C3) prevents us from boosting the performance of representation models with language-specific pre-processing like stemming, lemmatization and part-of-speech tagging, as is typically done in the literature \cite{alvarez2016topic, chen2012collaborative, DBLP:conf/webi/ZhaoT14}. Instead, our experimental methodology is language-agnostic.
%ingests
%noise into all representation models, due to the language-agnostic settings
%of our experiments. 
%given that language-specific techniques lie out of the scope
%of this work.

\textbf{Performance Measures.} We assess the
\underline{effectiveness} of representation models using the \textit{Average Precision}
($\mathbf{AP}$) of a user model $UM_s(u)$, which is the average of the Precision-at-n ($P@n$)  values for all
retweets. Formally \cite{chen2012collaborative,lu2012twitter}:
$AP(UM_s(u)) =1/|R(u)|\cdot\sum_{n=1}^{N}$$P@n\cdot RT(n)$,
where $P@n$ is the proportion of the top-n ranked tweets that
have been retweeted, $RT(n)$=1 if $n$ is a retweet and 0 otherwise, $N$ is the
size of the test set, and $|R(u)|$ is the total number of retweets in the
test set. Thus, $AP$ expresses the performance of a representation model over an
individual user. To calculate the performance of a user group $U$,
we define \textit{Mean Average Precision} (\textbf{MAP}) 
as the average $AP$ over all users in $U$.

To assess the \underline{robustness} of a representation model with respect to its
internal configuration, we consider its \textit{MAP deviation}, i.e., the difference
between the maximum and the minimum MAP of the considered
parameter configurations over a specific group of users. The lower the
MAP deviation, the 
%indicate 
higher the robustness.

To estimate the \underline{time efficiency} of representation models, we
employ two measures: \emph{(i)} The \textit{training time} ($\mathbf{TTime}$) captures 
the aggregated modeling time that is required for modeling all 60 users.
For topic models, this also includes the time that is required for training once
the model $M(s)$ from the entire train set.
\emph{(ii)} The \textit{testing time} ($\mathbf{ETime}$) expresses the total time that is required
for processing the test set of all 60 users, i.e., to compare all user models
with their testing tweets and to rank the latter in descending order of
similarity. 
%Note that 
For a fair comparison, we do not consider works that parallelize representation~models~(e.g.,~\cite{hierarchicalTopicModels,
DBLP:journals/pvldb/SmolaN10,DBLP:journals/pvldb/YuCZS17}), as most models are not adapted to distributed processing.

\begin{table}[!t]\centering
\caption{Configurations of the 5 context-agnostic (topic) models.
\textsf{CS}, \textsf{NP}, \textsf{UP} and \textsf{HP} stand for cosine similarity, no pooling, user pooling and hashtag pooling, respectively.
}
\vspace{-5pt}
{\small
{\renewcommand{\arraystretch}{1.2}
\begin{tabular}{ |l|| c | c | c | c | c |}
    \cline{2-6}
   	\multicolumn{1}{c|}{} & \textbf{\textsf{LDA}} & \textbf{\textsf{LLDA}} & 
   	\textbf{\textsf{BTM}} & \textbf{\textsf{HDP}} & \textbf{\textsf{HLDA}} \\ 
   	\hline
   	\hline
	\#Topics & \multicolumn{3}{c|}{\{50,100,150,200\}}& - & - \\
	\hline
	\#Iterations & \multicolumn{2}{c|}{\{1,000, 2,000\}} &
	\multicolumn{3}{c|}{1,000}
	\\
	\hline
	Pooling & \multicolumn{4}{c|}{\{\textsf{NP}, \textsf{UP}, \textsf{HP}\}} &
	\textsf{UP} \\
	\hline 
	$\alpha$ & \multicolumn{3}{c|}{50/\#Topics} & 1.0 & \{10, 20\} \\
	\hline
	$\beta$ & \multicolumn{3}{c|}{0.01} & \multicolumn{2}{c|}{\{0.1, 0.5\}} \\\hline
	$\gamma$ & - & - & - & 1.0 & \{0.5, 1.0\} \\
	\hline
	Aggregation function & \multicolumn{5}{c|}{ \{centroid, Rocchio\}}\\
	\hline
	Similarity measure& \multicolumn{5}{c|}{\textsf{CS}}\\
	\hline
	Fixed parameters & - & - & $r$=30 & - & levels=3 \\
	\hline
	\hline  
	\#Configurations & 48 & 48 & 24 & 12 & 16 \\
	\hline
\end{tabular}
}
}
%s\vspace{-7pt}
\label{tb:contextAgnosticConf}
\vspace{-15pt}
\end{table}

\textbf{Parameter Tuning.}
For each representation model, we tried a wide range of meaningful parameter
configurations. They are reported in Tables \ref{tb:contextAgnosticConf} and
\ref{tb:contextBasedConf}.
In total, we employed 223 different configurations -- excluding
%we ignored 
those 
%configurations that 
violating the \textit{memory constraint}
(the memory consumption should be less than 32GB RAM), or the \textit{time constraint} 
($TTime$ should be less than 5 days).
As a result, we excluded \textsf{PLSA} from our analysis, since it
violated the memory constraint for all configurations we considered. 

In total, 9
representation models were tested in our experiments.
For \textsf{LDA}, \textsf{LLDA} and \textsf{BTM}, $\alpha$ and $\beta$ were
tuned according to \cite{steyvers2007probabilistic}. For \textsf{LLDA}, the
number of topics refers to the latent topics assigned to every tweet in addition
to the tweet-specific labels. For the latter, we followed
\cite{ramage2010characterizing}, using:
%(except for the label \textit{@reply}, which is absent from our dataset):
\emph{(i)} one label for every hashtag that occurs more than 30 times in the
training tweets, \emph{(ii)} the question mark, \emph{(iii)} 9 categories of
emoticons (i.e., ``smile'', ``frown'', ``wink'', ``big grin'', ``heart'', ``surprise'',
``awkward'' and ``confused''), and \emph{(iv)} the @user label for the training tweets that mention a user as the first word.
%  does not contain reply information.
Most of these labels were quite frequent in our corpus and, thus, we considered
10 variations for them, based on \cite{ramage2010characterizing}.
For example, ``frown''\ produced the labels: :(-0 to :(-9. The only labels with no
variations are the hashtag ones and the emoticons ``big grin'', ``heart'',
``surprise'' and ``confused''. 
%Note that we ignored the .
% For LLDA, the number of topics refer to the latent
% labels, applied to every tweet, introduced in \cite{ramage2010characterizing}.

For \textsf{HLDA}, we did not employ the pooling strategies \textsf{NP} and
\textsf{HP} and more than three hierarchical levels, as these
configurations violated the time constraint. Hence, we only varied~$\alpha$~and~$\gamma$.

For \textsf{BTM}, we selected
1,000 iterations following \cite{yan2013biterm}. For individual tweets, we set 
the context window ($r$), i.e., the maximum distance between two words in a
biterm, equal to the size of the tweet itself. For the large pseudo-documents in
user and hashtag pooling, we set $r$=30 based on \cite{yan2013biterm}; for greater values, \textsf{BTM}
conveys no significant improvement over \textsf{LDA}, since the larger the
distance between the words in a biterm is, the more irrelevant are their topics. 
%Also, the 
Larger window sizes yield higher $TTime$, too.
% , degrading time efficiency. 

For bag models, some configuration combinations are invalid:
\textsf{JS} is applied only with \textsf{BF} weights, \textsf{GJS} only with
\textsf{TF} and \textsf{TF-IDF}, and the character-based n-grams, \textsf{CN},
are not combined with \textsf{TF-IDF}. Also, \textsf{BF} is exclusively coupled
with the sum aggregation function, which in this case is equivalent to applying
the boolean operator \textsf{OR} among the individual document models.
For the Rocchio algorithm, we set $\alpha$=0.8 and $\beta$=0.2,
and we used only the \textsf{CS} similarity measure in combination with the
\textsf{TF} and \textsf{TF-IDF} weights for those representation
sources that contain both positive and negative examples, namely \textsf{C},
\textsf{E}, \textsf{TE}, \textsf{RE},~\textsf{TC},~\textsf{RC}~and~\textsf{EF}.

\begin{table}[!t]\centering
\caption{Configurations of the 4 context-based models.
Remember that \textsf{BF} and \textsf{TF} stand for boolean and term frequency
weights, respectively, while (\textsf{G})\textsf{JS}, \textsf{CoS}, \textsf{VS}, and
\textsf{NS} denote the (generalized) Jaccard, the containment,
the value and the normalized value graph similarities, resp.}
\vspace{-5pt}
{\small
{\renewcommand{\arraystretch}{1.2}
\begin{tabular}{ |l|| c | c | c | c |}
    \cline{2-5}
   	\multicolumn{1}{c|}{} & \textbf{\textsf{TN}} & \textbf{\textsf{CN}} & 
   	\textbf{\textsf{TNG}} & \textbf{\textsf{CNG}} \\ 
   	\hline
   	\hline
	$n$ & \{1,2,3\} & \{2,3,4\} & \{1,2,3\} & \{2,3,4\} \\
	\hline
	Weighting scheme & \{\textsf{BF},\textsf{TF},\textsf{TF-IDF}\} &
	\{\textsf{BF},\textsf{TF}\} & - & - \\
	\hline
	Aggregation function & \multicolumn{2}{c|}{\{sum,  centroid, Rocchio\}} & - &
	- \\
	\hline
	Similarity measure & \multicolumn{2}{c|}{\{\textsf{CS}, \textsf{JS},
	\textsf{GJS}\}} & \multicolumn{2}{c|}{\{\textsf{CoS}, \textsf{VS}, \textsf{NS}\}} \\
	\hline
	\hline  
	\#Configurations & 36* & 21* & 9 & 9 \\
	\hline
	\multicolumn{5}{l}{* excluding invalid configuration combinations}
\end{tabular}
}
}
%\vspace{-12pt}
\label{tb:contextBasedConf}
\vspace{-15pt}
\end{table}
%\vspace{-8pt}
\section{Experimental Analysis}
\label{sec:analysis}
\vspace{-2pt}

\begin{figure*}[!t]
\vspace{-0.3in}
\centering
\includegraphics[width=0.89\textwidth]{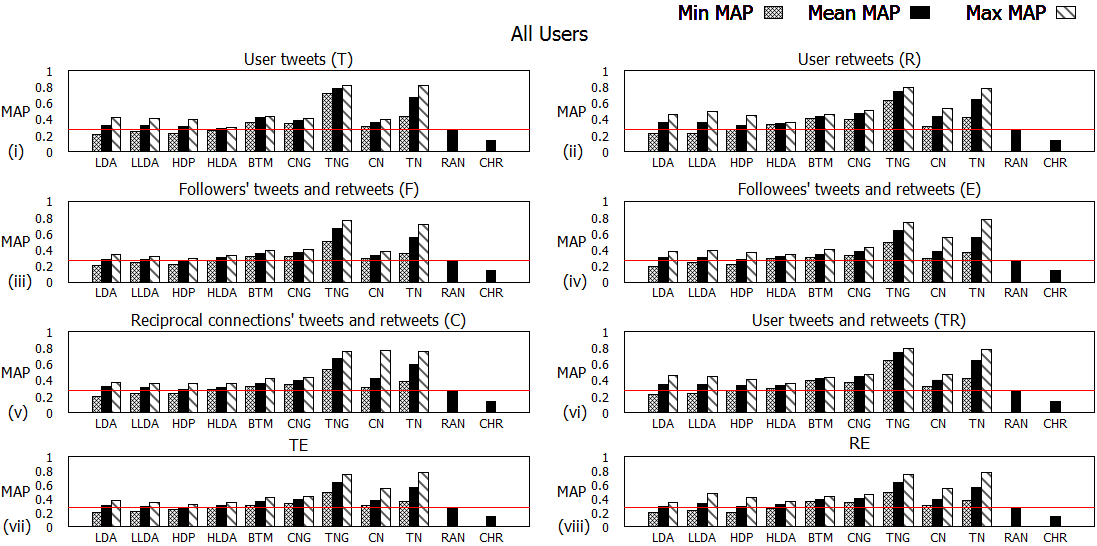}
\includegraphics[width=0.89\textwidth]{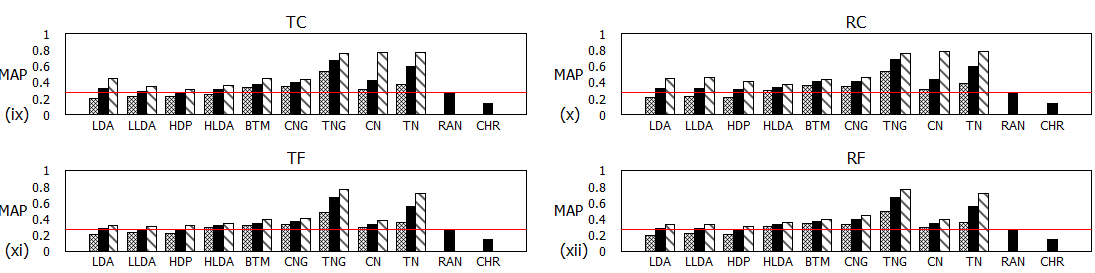}
\vspace{-9pt}
\caption{Effectiveness of the 9 representation models over \textsf{All
Users} in combination with the 5 individual representation sources and their 3
best performing pairwise combinations with respect to MAP.
Higher bars indicate better performance. The red line corresponds to the
performance of the best baseline, \textsf{RAN}.}
\label{fig:all}
\includegraphics[width=0.89\textwidth]{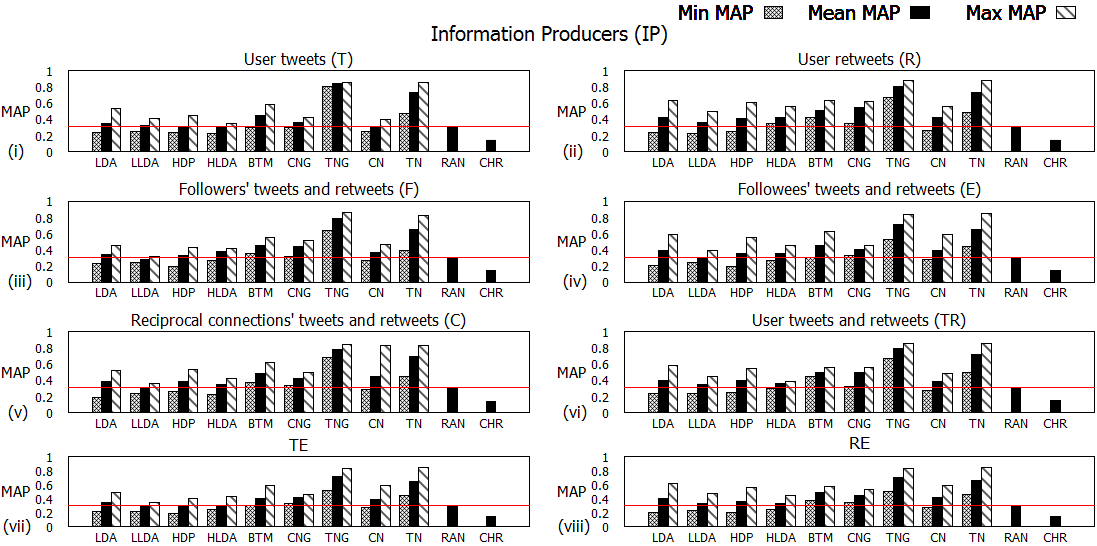}
\includegraphics[width=0.89\textwidth]{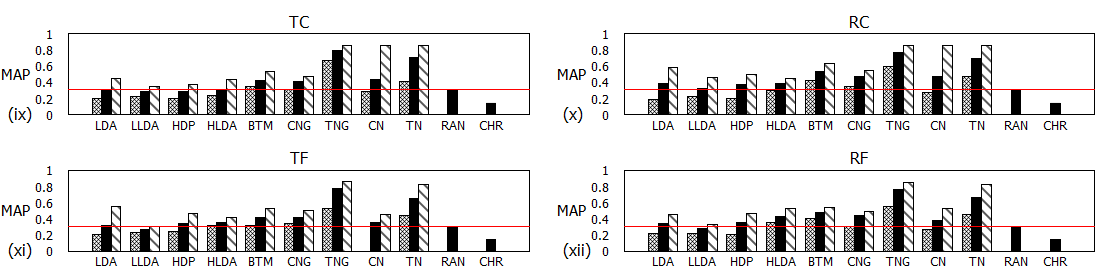}
\vspace{-10pt}
\caption{MAP values for the 9 representation models over the
\textsf{IP} users and the same 8 representation sources.}
\vspace{-15pt}
\label{fig:pip}
\end{figure*}

\begin{figure*}[!t]
\vspace{-0.3in}
\centering
\includegraphics[width=0.89\textwidth]{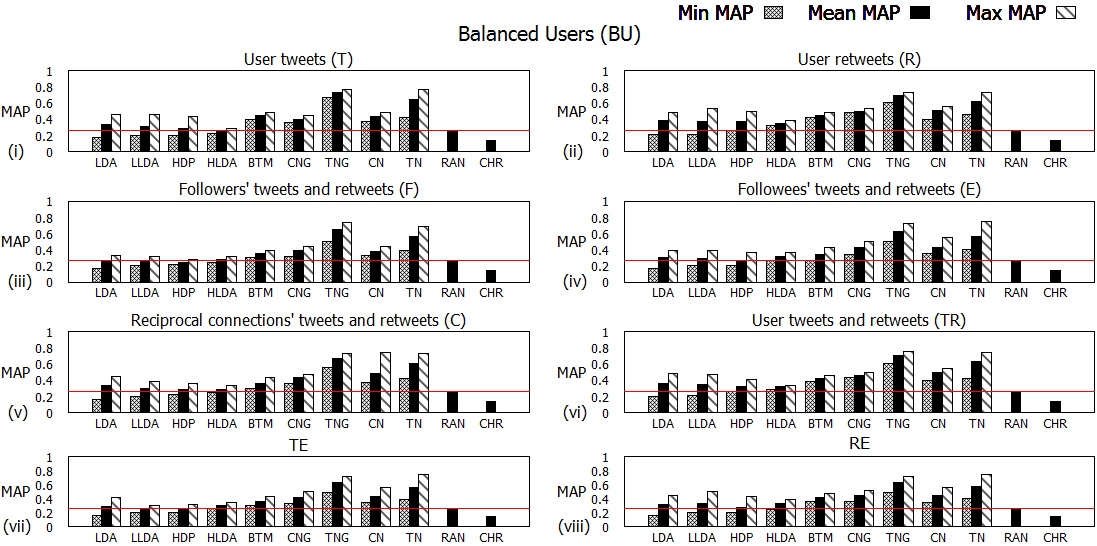}
\includegraphics[width=0.89\textwidth]{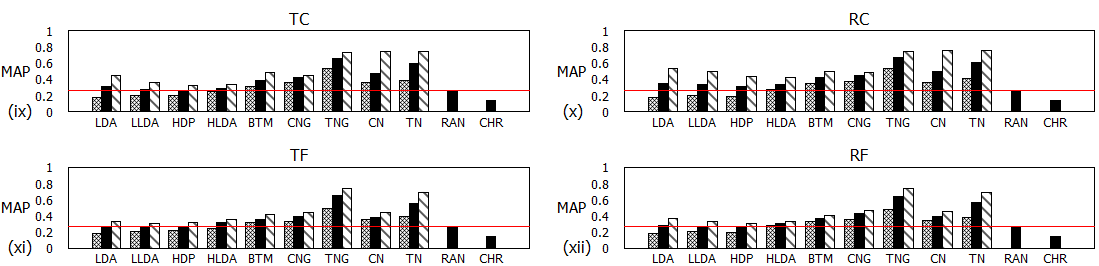}
\vspace{-9pt}
\caption{Effectiveness of the 9 representation models over the
\textsf{BU} users in combination with the 5 individual representation sources
and their 3 best performing pairwise combinations with respect to MAP.
Higher bars indicate better performance. The red line corresponds to the
performance of the best baseline, \textsf{RAN}.}
\label{fig:bu}
\includegraphics[width=0.89\textwidth]{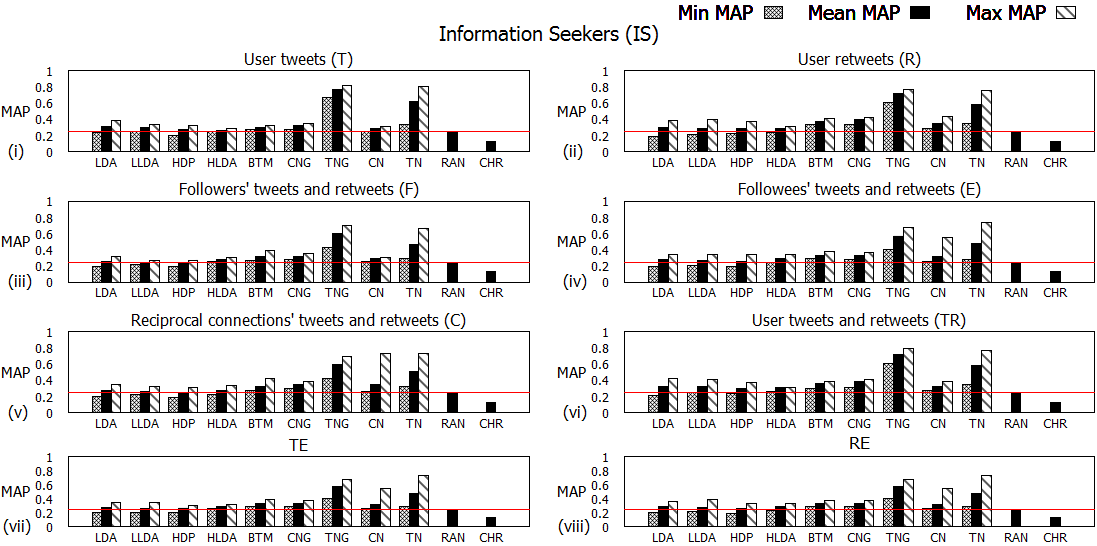}
\includegraphics[width=0.89\textwidth]{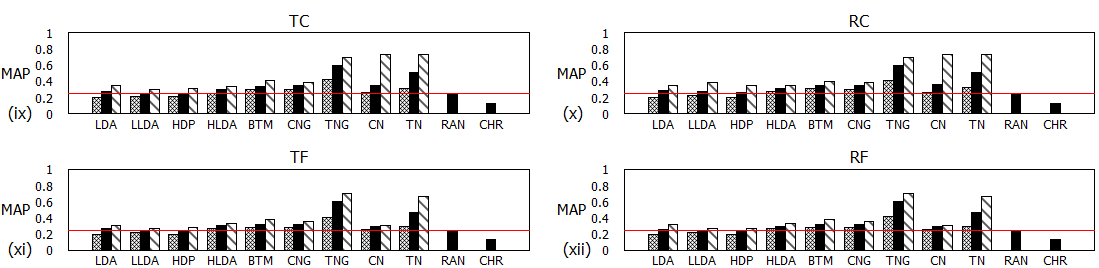}
\vspace{-10pt}
\caption{MAP values for the 9 representation models over the
\textsf{IS} users and the same 8 representation sources.}
\vspace{-15pt}
\label{fig:is}
\end{figure*}

\textbf{Effectiveness \& Robustness.} To assess the performance of the 9
representation models, we measure their \textit{Mean
MAP}, \textit{Min MAP} and \textit{Max MAP}, i.e., their average, minimum and
maximum MAP, respectively, over all relevant configurations for a particular
combination of a user type and a representation source.
The outcomes are presented in Figures \ref{fig:all}, \ref{fig:pip}, \ref{fig:bu} and \ref{fig:is} for \textsf{All Users}, \textsf{IP},
\textsf{BU} and \textsf{IS}, respectively. In every figure, we consider the five individual
representation sources (\textsf{T}, \textsf{R}, \textsf{F}, \textsf{E},
\textsf{C}) along with the eight pairwise combinations.
%that achieve the best average
%performance for each user type; these are \textsf{TR}, \textsf{RC} and
%\textsf{RE} across all user types. Due to space limitations, the remaining
%5 combinations are only presented in the extended version of the paper \cite{technicalReport}. Nevertheless, we consider them in the discussion
%of our experimental results.

Every diagram also reports the MAP for two
baseline methods:
(i) the \textit{Chronological Ordering} (\textsf{CHR}), which ranks the test
set from the latest tweet (first position) to the earliest one (last position), 
and (ii) the \textit{Random Ordering} (\textsf{RAN}), which sorts the test set
in an arbitrary order. For \textsf{RAN}, we performed 1,000 iterations per
user and considered the overall average per user type.

Starting with \textsf{All Users} in Figure \ref{fig:all}, we observe that
\textsf{TNG} consistently outperforms all other models across all representation sources.
Its Mean MAP fluctuates between 0.625 (\textsf{EF}) and 0.784 (\textsf{T}).
The second most effective model is \textsf{TN}, whose Mean MAP ranges
from 0.541 (\textsf{EF}) to 0.673 (\textsf{T}).
The dominance of \textsf{TNG} over \textsf{TN} is statistically
significant ($p$$<$$0.05$) and should be attributed to its ability to capture
the relations between neighboring token n-grams through the weighted edges that
connect them. In this way, \textsf{TNG} incorporates global contextual
information into its model and inherently alleviates sparsity (Challenge C1). Instead,
\textsf{TN} exclusively captures local contextual information in the form of token
sequences. The same holds for Challenges C2 and C4: both models fail
to identify misspelled or non-standard token n-grams, but \textsf{TNG} is better
suited to capture their patterns by encapsulating their neighborhood. 

\begin{table*}[h]\centering
\caption{Performance of all 13 representation sources over the 4
user types with respect to Min(imum), Mean and Max(imum) MAP across all configurations
of the 9 representation models. The rightmost column presents the average
performance per user type.}
\vspace{-8pt}
{\fontsize{7.2pt}{7pt}\selectfont
{\setlength\extrarowheight{1.2pt}
\begin{tabular}{ |l l|| c  c  c  c  c || c  c  c  c  c   c  c  c || c |}
    \cline{3-16} 
	\multicolumn{2}{c|}{} & \textbf{\textsf{R}} &\textbf{\textsf{T}}
	&\textbf{\textsf{E}} &\textbf{\textsf{F}} &\textbf{\textsf{C}}
	&\textbf{\textsf{TR}} &\textbf{\textsf{RE}} &\textbf{\textsf{RF}}
	&\textbf{\textsf{RC}} &\textbf{\textsf{TE}} &\textbf{\textsf{TF}}
	&\textbf{\textsf{TC}} &\textbf{\textsf{EF}} & Average\\
	\hline
	\hline
	 & Min MAP	&	0.225	&	0.217	&	0.196	&	0.208	&	0.199	&	0.222	&	0.205	&	0.199	&
	 0.212	&	0.201	&	0.205	&	0.201	&	0.198	& 0.207 \\
	 \textbf{\textsf{All Users}} & Mean MAP	&	0.456	&	0.429	&	0.392	&
	 0.378	&	0.410	&	0.448	&	0.402	&	0.386	&	0.427	&	0.387	&	0.377	&	0.406	&	0.377
	 &	0.406 \\ 
	 & Max MAP	&	0.796	&	0.816	&	0.771	&	0.764	&	0.768	&	0.797	&	0.775	&
	 0.758	&	0.783	&	0.775	&	0.764	&	0.776	&	0.780	& 0.779\\
\hline
\hline

 & Min MAP	&	0.193	&	0.199	&	0.192	&	0.191	&	0.196	&	0.212	&	0.199	&	0.195	&
 0.188 &	0.200	&	0.192	&	0.195	&	0.195	&	0.196
 \\ 

\textbf{\textsf{IS}} & Mean MAP	&	0.415	&	0.383	&	0.351	&	0.336	&
0.364	&	0.407	&	0.357	&	0.340	&	0.374	&	0.347	&	0.342	&	0.357	&	0.336	&	0.362\\
 
& Max MAP	&	0.776	&	0.818	&	0.733	&	0.699	&	0.732	&	0.790	&	0.732	&	0.699	&
0.733 &	0.732	&	0.699	&	0.733	&	0.736	&	0.739
\\

\hline
\hline

& Min MAP	&	0.212	&	0.174	&	0.168	&	0.192	&	0.168	&	0.198	&	0.168	&	0.182	&
0.178 &	0.167	&	0.179	&	0.171	&	0.175	& 0.179
\\ 
\textbf{\textsf{BU}} & Mean MAP	&	0.474	&	0.430	&	0.400	&	0.376	&
0.420	&	0.453	&	0.419	&	0.389	&	0.443	&	0.391	&	0.381	&	0.406	&	0.376	&	0.412
\\
& Max MAP	&	0.733	&	0.775	&	0.747	&	0.746	&	0.745	&	0.753	&	0.746	&	0.741	&
0.753	&	0.746	&	0.741	&	0.743	&	0.741	& 0.747
\\

\hline
\hline

& Min MAP	&	0.243	&	0.220	&	0.200	&	0.198	&	0.186	&	0.242	&	0.200	&	0.202	&
0.182 &	0.189	&	0.212	&	0.198	&	0.208	& 0.206
\\ 

\textbf{\textsf{IP}} & Mean MAP	&	0.524	&	0.446	&	0.460	&	0.458	&
0.488	&	0.493	&	0.470	&	0.467	&	0.497	&	0.428	&	0.443	&	0.444	&	0.440	&	0.466
\\
& Max MAP	&	0.878	&	0.857	&	0.854	&	0.858	&	0.839	&	0.854	&	0.854	&	0.856	&
0.854	&	0.854	&	0.858	&	0.857	&	0.854	& 0.856
\\
\hline

\end{tabular}
}
}
%\vspace{-7pt}
\label{tb:sourcesAndTypes}
\vspace{-10pt}
\end{table*}

Regarding \textit{robustness}, \textsf{TN} is more sensitive to its
configuration. Its MAP deviation ranges from 0.359 
(\textsf{TF}, \textsf{RF}) to 0.444 (\textsf{EF}), while
for \textsf{TNG}, it fluctuates between 0.096
(\textsf{T}) and 0.281 (\textsf{EF}). In all cases, the difference
between the two models is statistically significant ($p$$<$$0.05$). \textsf{TNG}
is superior, because its performance is fine-tuned by just two parameters: the size of
n-grams ($n$) and the similarity measure. \textsf{TN} additionally involves
the aggregation function and the weighting scheme, increasing
drastically its possible configurations. In general, \textit{every
configuration parameter acts as a degree of freedom for a representation model;
the higher the overall number of parameters is, the more flexible the model gets and the less robust
it is expected to be with respect to its configuration}.

Comparing the character-based instantiations of bag and graph models, we notice
that their difference in Mean Map is statistically insignificant.
For \textsf{CNG}, it fluctuates between 0.368 (\textsf{TF}) and 0.477
(\textsf{R}), while for \textsf{CN}, it ranges from 0.334 (\textsf{TF})
to 0.436 (\textsf{RC}). This implies that \textit{there is no benefit in
considering global contextual information for representation models
that are based on character n-grams. The strength of these models lies
in the local contextual information that is captured in the sequence of
characters}.

Regarding robustness, the relative sensitivity of character-based models
exhibits the same patterns as their token-based counterparts: the bag models are
significantly less robust than the graph ones, due to their larger number of
parameters and configurations. In more detail, the MAP deviation
ranges from 0.061 (\textsf{T}) to 0.114 (\textsf{RF}) for \textsf{CNG} and from
0.077 (\textsf{T}) to 0.476~(\textsf{RC})~for~\textsf{CN}.

Concerning the topic models, we observe that \textsf{BTM} consistently achieves
the highest effectiveness across all representation sources. Its Mean MAP fluctuates between 0.340 (\textsf{EF}) and 0.434 (\textsf{R}).
All other topic models exhibit practically equivalent
performance: their highest Mean MAP ranges from 0.337 (for
\textsf{HDP} with \textsf{TR}) to 0.360 (for \textsf{LLDA} with \textsf{R}), whereas their lowest Mean MAP fluctuates
between 0.265 (for \textsf{HDP} over \textsf{F}) and 0.270 (for \textsf{LLDA}
over \textsf{TF}). As a result, \textsf{HDP}, \textsf{LDA}, \textsf{LLDA} and
\textsf{HLDA} outperform only \textsf{CHR} to a large extent. This indicates
that \textit{recency alone constitutes an inadequate criterion for recommending content in
microblogging platforms. Any model that considers the personal
preferences of a user offers more accurate suggestions}.

Compared to the second baseline (\textsf{RAN}), \textsf{HDP}, \textsf{LDA}, \textsf{LLDA} and
\textsf{HLDA} are more effective, but to a minor extent. The Mean
MAP of \textsf{RAN} amounts to 0.270,
thus, some configurations of these topic models perform consistently worse than
\textsf{RAN}, across all representation sources. Most of these configurations correspond to
the absence of pooling (\textsf{NP}), where every tweet is considered as
an individual document. In these settings, these four topic models fail to
extract distinctive patterns from any representation source,
producing noisy user and document models. This suggests that \textit{sparsity is the main obstacle to most topic models}.

Regarding robustness, we can distinguish the 5 topic models into two categories:
the first one involves 3 models that are highly sensitive to their
configuration, namely \textsf{LDA}, \textsf{LLDA} and \textsf{HDP}.
Their MAP deviation starts from 0.119, 0.075, and 0.077, respectively, and
raises up to 0.250, 0.264 and 0.211, respectively. These values are extremely
high, when compared to their absolute Mean MAP. This means that 
\textit{extensive fine-tuning is required for successfully applying most topic models to text-based PMR}. In contrast, MAP deviation fluctuates between 0.034 and 0.109
for both \textsf{HLDA} and \textsf{BTM}. For the former, this is probably caused by the
limited number of configurations that satisfied our time constraint, while for the latter, it should be attributed to its Twitter-specific functionality.

\textbf{Discussion.} We now discuss the novel insights that can be deduced from our experimental analysis. We start by comparing token- with character-based models. We observe that the former
are significantly more effective than the latter for both bag and graph models. 
At first, this seems counter-intuitive, as \textsf{CNG} and \textsf{CN} are in a
better position to address Challenge C1: they extract more
features from sparse documents than \textsf{TNG} and \textsf{TN},
respectively~\cite{DBLP:journals/www/0001GP16}. They are also better
equipped to address Challenges C2 and C4:
by operating at a finer granularity, they can identify similarities even
between noisy and non-standard tokens. Yet, the character-based models seem to
accumulate noise when aggregating individual tweet models into a user
model. Their similarity measures fail to capture distinctive
information about the real interests of a user, yielding high scores for
many irrelevant, unseen documents. The lower $n$ is, the more intensive is this
phenomenon. In fact, most bigrams are shared by both relevant and irrelevant
examples, which explains why the poorest performance corresponds to $n$=2 for both
\textsf{CNG} and \textsf{CN}. 
The only advantage of character-based models over their token-based counterparts
is their higher robustness. However, their lower values for MAP deviation are
probably caused by their lower absolute values for MAP.

Among the topic models, \textsf{BTM} consistently performs best with
respect to effectiveness and robustness. Its superiority 
%of \textsf{BTM} over the other topic models   
stems from the two inherent
characteristics that optimize its functionality for the short and noisy documents in Twitter:
\emph{(i)} it considers pairs of words (i.e., biterms), instead of individual
tokens, and \emph{(ii)} it bypasses sparsity (Challenge C1) by capturing topic
patterns at the level of entire corpora, instead of extracting them from
individual documents.
Compared to context-based models, though, \textsf{BTM} is significantly
less effective than the token-based bag and graph models. Its performance
is very close to the character-based models, especially
\textsf{CNG}, with their difference in terms of effectiveness and
robustness being statistically insignificant. 

Generally, we can conclude that all topic models
fail to improve the token-based bag and graph models in the context of Content-based PMR. Both \textsf{TN} and \textsf{TNG} achieve twice as high
Mean MAP, on average, across all representation sources.
The poor performance of topic models is caused by two factors: \emph{(i)} they
disregard the contextual information that is encapsulated in word ordering, and
\emph{(ii)} the Challenges C1 to C4 of Twitter. Indeed, most topic models were
originally developed to extract topics from word co-occurrences in individual
documents, but the sparse and noisy co-occurrence patterns in the short text of
tweets reduce drastically their effectiveness.

\begin{figure*}[t!]\centering
\includegraphics[width=17cm,keepaspectratio]{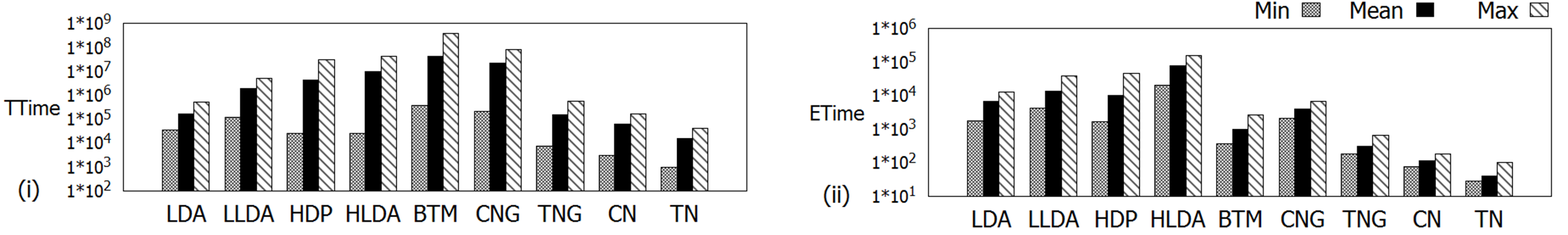}
\vspace{-8pt}
\caption{Time efficiency of the 9 representation models with respect to
\emph{(i)} Training Time ($TTime$), and \emph{(ii)} Testing Time ($ETime$), on
average across all configurations, representation sources and users.
Lower bars indicate better performance.}
\label{fig:time}
\vspace{-15pt}
\end{figure*}

\textbf{User Types.} We notice that the \textit{relative} performance of
representation models in Figures~\ref{fig:pip} (\textsf{IP}), \ref{fig:bu} (\textsf{BU}) and \ref{fig:is} (\textsf{IS}) remains
practically the same as in Figure~\ref{fig:all} (\textsf{All Users}). Yet, there are significant
differences in their \textit{absolute} performance. On average, across all models and
representation sources, \textsf{IP} users have higher Mean MAP than \textsf{IS}
and \textsf{BU} users by 30\% and 13\% respectively, while \textsf{BU} surpasses
\textsf{IS} by 15\%. 
Compared to \textsf{All Users}, \textsf{IP} increases Mean MAP
by 12\%, \textsf{BU} by just 1\%, while \textsf{IS}
decreases it by 11\%, on average, across all models and
representation sources. These patterns are also demonstrated in the
rightmost column of Table~\ref{tb:sourcesAndTypes}, which essentially
expresses the average values for Min(imum), Mean and Max(imum) MAP per user type
across all combinations of representation models, configurations and
representation sources.

Therefore, we can conclude that  \textit{the more information an
individual user produces, the more reliable are the models that represent her
interests}, and vice versa: \textit{taciturn users are the most difficult to model}. 
This should be expected for \textsf{R} and \textsf{T},
since the posting activity of a user increases the content that is available for
building their models. However, this pattern applies to the
other representation sources, too, because \textsf{IP} users are
typically in closer contact with their followees, followers and reciprocal users, who thus
can represent their interests effectively.
The opposite holds for \textsf{IS} users. In the middle of these two
extremes lie \textsf{BU} users, which exhibit a balanced activity in all respects. 

Overall, these patterns suggest that the user categories we defined in Section \ref{sec:preliminaries} have a real impact on the performance of Content-based PMR. Therefore, they should be taken into account when designing a Content-based PMR approach, as different user types call for different recommendation methods.

\textbf{Representation Sources.} We now examine the relative effectiveness of
the five representation sources and their eight pairwise combinations. 
Table \ref{tb:sourcesAndTypes} reports the performance for every 
combination of a user type and a representation source with respect to
Min(imum), Mean and Max(imum) MAP over all configurations of the nine
representation models. %The results are grouped by~user~type.

Starting with individual representation sources, we observe that
\textsf{R} consistently achieves the highest Mean MAP across all user types.
This suggests that \textit{retweets constitute the most effective means of capturing
user preferences under all settings}, as users repost tweets that
have attracted their attention and approval.

The second best individual source is \textsf{T}. This applies to all user types,
except for \textsf{IP}, where \textsf{T} actually exhibits the worst performance
across all individual sources. These patterns suggest that \textit{the
tweets of a user offer valuable information for her interests as long as her posting ratio is lower
than 2}; such users post their messages thoughtfully, when they have something
important to say. Instead, the \textsf{IP} users are hyperactive, posting quite
frequently careless and noisy messages that do not reflect their
preferences, e.g., by engaging into irrelevant discussions with other users.

Among the remaining individual sources, \textsf{C} achieves the best
performance, followed by \textsf{E} and \textsf{F}. 
This pattern is consistent across all user types, with the
differences being statistically significant in most cases. We can deduce,
therefore, that \textit{the reciprocal connections in Twitter reflect friendships
between users that share common interests to a large extent}. Instead, the
one-way connections offer weaker indications of common tastes among users,
especially when they are not initiated by the ego user: the
followers' posts (\textsf{F}) invariably result in noisy user models.
 
For the eight pairs of sources, we observe the following patterns: \emph{(i)} All
combinations of \textsf{R} with another source \textsf{X} result in higher performance for \textsf{X}, with the difference being statistically significant. Hence, \textsf{R} is able to
enrich any representation source with valuable information about user
preferences.
\emph{(ii)} All
combinations of \textsf{T} with another source result in a (usually
insignificant) lower performance in all cases but two, namely \textsf{TF} over
\textsf{IS} and \textsf{BU}. This means that \textsf{T} typically conveys noisy, irrelevant information.
\emph{(iii)} For both \textsf{R} and \textsf{T}, all pairwise combinations  
degrade their own performance. The only exception is \textsf{TR}, which improves the effectiveness of \textsf{T}.

On the whole, we can conclude that \textit{a user's retweets (\textsf{R}) should be used as the
sole information source for building her model}. There is no need to combine it with \textsf{T}.

\textbf{Time Efficiency.}
Figures \ref{fig:time}(i) and (ii) depict the minimum, average and maximum
values for $TTime$ and $ETime$, respectively, for every representation model
across all configurations, representation sources and users. The vertical axes are logarithmic, with lower
values indicating better performance.

Among the global context-aware models, we observe that on average, \textsf{TNG}
is faster than \textsf{CNG} with respect to both $ETime$ and $TTime$ by 1 and 2
orders of magnitude, respectively. Similarly, among the local context-aware
models, \textsf{TN} is faster than \textsf{CN} by 4 and 3 times, respectively.
The reason in both cases is the relative dimensionality of token- and 
character-based models. Typically, the latter yield a much larger feature
space than the former, depending, of course, on the size of the
n-grams -- the larger $n$ is, the more character n-grams are extracted from a
corpus~\cite{DBLP:journals/www/0001GP16}
%, with higher dimensionality resulting in 
and the more time-consuming is their processing.
  
Among the topic models, \textsf{BTM} exhibits the highest $TTime$. The reason is
that it operates on the level of biterms, which result in a much higher
dimensionality than the individual tokens that lie at the core of the other
topic models. However, our analysis does not consider the most time-consuming
configurations of the second worst model (\textsf{HLDA}), as they violated the
time constraint. In practice, \textsf{HLDA} is expected to be slower than
\textsf{BTM}, since its nonparametric nature lies in exploring the possible
L-level hierarchical trees during topic inference. On the other extreme,
\textsf{LDA} is the fastest topic model, while \textsf{LLDA} is almost 5
times faster than \textsf{HDP}, on average, due to its simpler models.

These patterns are slightly altered in the case of $ETime$. The worst
performance by far corresponds to \textsf{HLDA}, due to its nonparametric
functionality, while \textsf{BTM} turns out to be the fastest model.
Unlike the other topic models, which perform Gibbs sampling for topic inference,
\textsf{BTM} simply iterates over the biterms in a document $d$ in order to
calculate $P(z|d)$ for each topic $z \in Z$.

\begin{table*}\scriptsize\centering
	\caption{The most effective configuration per representation model and
	information source. Remember that \textsf{NP}, \textsf{UP} and \textsf{HP}
	stand for no, user and hashtag pooling, respectively; \textsf{BF} and
	\textsf{TF} denote boolean and term frequency weights, respectively, while
	\textsf{CS}, (\textsf{G})\textsf{JS}, \textsf{CoS}, \textsf{VS}, and
	\textsf{NS} stand for the cosine, (generalized) Jaccard,
	containment, value and normalized value similarities, respectively. \textsf{Cen.},
	\textsf{Ro.} and \textsf{Sum} represent the Centroid, Rocchio and Sum
	aggregation functions, respectively. Finally, the symbols SM, WS, AF, PS, \#I and \#T in the second column stand for Similarity Measure, Weighting Scheme, Aggregation Function, Pooling Scheme, number of iterations and number of topics, respectively.}
	\vspace{-5pt}
{\small
\renewcommand{\arraystretch}{1.1}
\setlength{\tabcolsep}{3.82pt}
	\begin{tabular}{ |l | l||c|c|c|c|c||c|c|c|c|c|c|c|c| }
		\cline{3-15}
		\multicolumn{2}{c|}{} & \textbf{\textsf{R}} & \textbf{\textsf{T}} &
		\textbf{\textsf{F}} & \textbf{\textsf{E}} & \textbf{\textsf{C}} &
		\textbf{\textsf{TR}} & \textbf{\textsf{TE}} & \textbf{\textsf{RE}} &
		\textbf{\textsf{TC}} & \textbf{\textsf{RC}} & \textbf{\textsf{TF}} &
		\textbf{\textsf{RF}} & \textbf{\textsf{EF}} \\
		\hline
		\hline
		\multirow{2}{*}{\textbf{\textsf{CNG}}} & $n$ & 4 & 4 & 4 & 4 & 4 & 4 & 4 & 4 &
		4 & 4 & 4 & 4 & 4 \\
		& SM & \textsf{NS} & \textsf{VS} &
		\textsf{CoS} & \textsf{CoS} & \textsf{CoS} &
		\textsf{CoS} & \textsf{CoS} & \textsf{CoS} &
		\textsf{CoS} & \textsf{CoS} & \textsf{CoS} &
		\textsf{CoS} & \textsf{CoS} \\
		\hline
		\hline
		\multirow{2}{*}{\textbf{\textsf{TNG}}} & $n$ & 3 & 3 & 3 & 3 & 3 & 3 & 3 & 3 &
		3 & 3 & 3 & 3 & 3 \\
		& SM & \textsf{VS} & \textsf{VS} &
		\textsf{VS} & \textsf{VS} & \textsf{VS} & \textsf{VS} & \textsf{VS} &
		\textsf{VS} & \textsf{VS} & \textsf{VS} & \textsf{VS} & \textsf{VS} &
		\textsf{VS} \\
		\hline
		\hline
		\multirow{4}{*}{\textbf{\textsf{CN}}} & $n$ & 4 & 4 & 4 & 4 & 3 & 4 & 4 & 4 &
		3 & 3 & 4 & 4& 4 \\
		& WS & \textsf{TF} & \textsf{TF} & \textsf{TF} & \textsf{TF} &
		\textsf{TF} & \textsf{TF} & \textsf{TF} & \textsf{TF} & \textsf{TF} &
		\textsf{TF} & \textsf{TF} & \textsf{TF} & \textsf{TF} \\
		& AF & Cen. & Sum & Cen. & Ro. & Ro. & Cen. & Ro. & Ro. & Ro. & Ro. &
		Cen. & Cen. & Ro. \\
		& SM & \textsf{CS} & \textsf{GJS} &
		\textsf{CS} & \textsf{CS} & \textsf{CS} & \textsf{CS} & \textsf{CS} &
		\textsf{CS} & \textsf{CS} & \textsf{CS} & \textsf{CS} & \textsf{CS} &
		\textsf{CS} \\
		\hline
		\hline
		\multirow{4}{*}{\textbf{\textsf{TN}}} & $n$ & 3 & 3 & 3 & 3 & 3 & 3 & 3 & 3 &
		3 & 1 & 3 & 3 & 3\\
		& WS & \textsf{BF} & \textsf{BF} & \textsf{TF-IDF} &
		\textsf{TF-IDF} & \textsf{TF-IDF} & \textsf{BF} & \textsf{TF-IDF} &
		\textsf{TF-IDF} & \textsf{TF-IDF} & \textsf{TF-IDF} & \textsf{TF-IDF} &
		\textsf{TF-IDF} & \textsf{TF-IDF}\\
		& AF & Sum & Sum & Cen. & Ro. & Ro. & Sum & Ro. & Ro. & Ro. & Ro. &
		Cen. & Cen. & Ro. \\
		& SM & \textsf{JS} & \textsf{JS} &
		\textsf{CS} & \textsf{CS} & \textsf{CS} & \textsf{JS} & \textsf{CS} &
		\textsf{CS} & \textsf{CS} & \textsf{CS} & \textsf{CS} & \textsf{CS} &
		\textsf{CS} \\
		\hline
		\hline
		\multirow{4}{*}{\textbf{\textsf{LDA}}} & PS & \textsf{UP} & \textsf{UP} &
		\textsf{HP} & \textsf{HP} & \textsf{UP} & \textsf{UP} & \textsf{UP} & 
		\textsf{HP} & \textsf{UP} & \textsf{UP} & \textsf{HP} & \textsf{HP} &
		\textsf{UP} \\
		& \#I & 2,000 & 2,000 & 1,000 & 2,000 & 1,000 & 1,000 & 1,000 & 1,000
		& 2,000 & 2,000 & 1,000 & 1,000 & 1,000 \\
		& \#T & 150 & 100 & 100 & 50 & 50 & 150 & 200 & 150 & 100 & 100 & 150 &
		150 & 150 \\
		& AF & Cen. & Cen. & Cen. & Ro. & Cen. & Cen. & Cen. & Cen. & Ro. & Ro. & Cen. & Cen.
		& Ro. \\
		\hline
		\hline
		\multirow{4}{*}{\textbf{\textsf{LLDA}}} & PS & \textsf{UP} & \textsf{UP}
		& \textsf{UP} & \textsf{UP} & \textsf{NP} & \textsf{UP} & \textsf{UP} &
		\textsf{UP} & \textsf{UP} & \textsf{UP} & \textsf{UP} & \textsf{UP} &
		\textsf{UP} \\
		& \#I & 1,000 & 2,000 & 2,000 & 2,000 & 1,000 & 2,000 & 1,000 & 2,000
		& 1,000 & 1,000 & 1,000 & 2,000 & 2,000 \\
		& \#T & 200 & 50 & 150 & 50 & 100 & 150 & 200 & 50 & 200 & 50 & 150 &
		200 & 200 \\
		& AF & Cen. & Cen. & Cen. & Ro. & Ro. & Cen. & Ro. & Ro. & Ro. & Ro. & Cen. & Cen. &
		Cen. \\
		\hline
		\hline
		\multirow{3}{*}{\textbf{\textsf{HDP}}} & PS & \textsf{UP} & \textsf{UP} &
		\textsf{HP} & \textsf{UP} & \textsf{HP} & \textsf{UP} & \textsf{UP} &
		\textsf{UP} & \textsf{HP} & \textsf{UP} & \textsf{UP} & \textsf{UP} &
		\textsf{UP} \\
		& $\beta$ & 0.1 & 0.1 & 0.1 & 0.1 & 0.1 & 0.1 & 0.1 & 0.1 & 0.1 & 0.1 & 0.1 &
		0.1 & 0.5 \\
		& AF & Cen. & Cen. & Cen. & Ro. & Ro. & Cen. & Ro. & Ro. & Cen. & Ro. & Cen. &
		Cen. & Cen. \\
		\hline
		\hline
		\multirow{4}{*}{\textbf{\textsf{HLDA}}} & $\alpha$ & 10 & 10 & 10 & 20 & 10 &
		20 & 20 & 10 & 10 & 20 & 20 & 20 & 10 \\
		& $\beta$ & 0.1 & 0.1 & 0.1 & 0.5 & 0.1 & 0.1 & 0.1 & 0.1 & 0.1 & 0.1 & 0.1
		& 0.1 & 0.1 \\
		& $\gamma$ & 1.0 & 0.5 & 1.0 & 0.5 & 1.0 & 1.0 & 1.0 & 1.0 & 1.0 & 1.0 & 1.0
		& 0.5 & 0.5 \\
		& AF & Cen. & Cen. & Cen. & Ro. & Cen. & Cen. & Cen. & Ro. & Cen. & Ro. & Cen. & Cen.
		& Cen. \\
		\hline
		\hline
		\multirow{3}{*}{\textbf{\textsf{BTM}}} & PS & \textsf{NP} & \textsf{NP} &
		\textsf{UP} & \textsf{UP} & \textsf{UP} & \textsf{UP} & \textsf{UP} &
		\textsf{UP} & \textsf{UP} & \textsf{HP} & \textsf{UP} & \textsf{UP} &
		\textsf{UP}\\
		& \#T & 150 & 200 & 200 & 150 & 150 & 150 & 200 & 150 & 100 & 200 & 150
		& 50 & 100 \\
		& AF & Cen. & Cen. & Cen. & Cen. & Cen. & Cen. & Cen. & Cen. & Cen. & Ro. & Cen. &
		Cen. & Cen. \\
		\hline
	\end{tabular}
	\label{configs-tb}
	\vspace{-8pt}
}
\end{table*}

Comparing the model categories between them, we observe that the graph models
are more time-consuming than their bag counterparts: on average, \textsf{TNG}
and \textsf{CNG} are 1 and 2 orders of magnitude slower than \textsf{TN} and 
\textsf{CN}, respectively, for both $TTime$ and $ETime$. This should be
attributed to the contextual information they incorporate in their edges, whose
number is much larger in the case of \textsf{CNG}. Similarly, most topic models are at least 1 order of magnitude slower than
their base model, \textsf{TN}, for both time measures, due to the time required
for topic inference.
Overall, \textit{\textsf{TN} is consistently the most efficient representation
model, due to the sparsity of tweets and the resulting low dimensionality.}

These patterns are consistent across all user types and representation sources.
For the latter, we also observed that the size of the train set 
affects directly $TTime$ for all representation models: the more
tweets are used to build a user model, the more patterns are extracted, degrading time efficiency.

\textbf{Configurations.}
Another way of assessing the robustness of representation models is to examine
the stability of their best performing configurations under various settings. To
this end, Table \ref{configs-tb} presents for every model the
configuration that achieves the highest Mean MAP for each representation
source, on average, across all user types. The more consistent its configuration is, the more robust is the model and the easier it is to fine-tune it.

Starting with the global context-aware models, we observe that for \textsf{TNG},
the same configuration achieves the best performance in all cases: token
tri-gram graphs ($n$=3) in combination with \textsf{VS}. The
configurations of \textsf{CNG} are quite robust, too, with character four-gram
graphs ($n$=4) consistently performing the best. In this case, though, the
similarity measure plays a minor role, as the difference in Mean MAP between the
best and the worst one is less than 0.005. Yet, \textsf{CoS} achieves the
highest Mean MAP for all sources, except \textsf{R} and \textsf{T}. These
patterns suggest that \textit{the graph models can be easily configured} and
that \textit{the longest n-grams tend to capture more effectively the distinguishing
patterns in microblogs}, at least for the considered values.
%$n{\in}[2,4]$ (for higher values, \textsf{CNG} ends up using entire tokens, thus behaving like \textsf{TNG}). 
This should be attributed to the larger neighborhood they examine for co-occurrence patterns, since their window size is also $n$.

Similar patterns apply to the local context-aware models, as the best
configurations of \textsf{TN} and \textsf{CN} are consistently dominated by the
longest n-grams: $n$=3 and $n$=4, respectively. The bag models are also robust with respect to their weighting
schemes and similarity measures. For \textsf{CN}, the best performance is
achieved by \textsf{TF} in combination with \textsf{CS} across all
sources, but one (\textsf{T}). In 10 out of 13 cases, the same similarity
measure works best for \textsf{TN}, but in conjunction with \textsf{TF-IDF} weights.
Regarding the aggregation strategy, Rocchio outperforms Sum and
Centroid for all representation sources that contain both negative
and positive tweets, namely \textsf{E}, \textsf{C}, \textsf{TE},
\textsf{RE}, \textsf{EF}, \textsf{TC}, \textsf{RC} and \textsf{EF}. In the other cases,
there is no clear winner between Sum and Centroid.

Regarding the topic models, we observe that they are highly sensitive to the
configuration of most parameters. This suggests that \textit{their
fine-tuning is a non-trivial task, depending heavily on the data at hand}
(e.g., the representation source). There are only few exceptions to this rule:
$\beta$=0.1 in all cases but one for \textsf{HDP} and \textsf{HLDA}, while
\textsf{UP} is the top performing pooling strategy in the vast majority of
cases. Indeed, \textsf{HP} and \textsf{NP} exhibit higher effectiveness in
just 9 and 3 out of 65 combinations of topic models and representation sources,
respectively. This should be expected for \textsf{NP}, where every tweet
corresponds to an individual document and the topic models are applied to
sparse documents. That's why \textsf{NP} achieves high scores only with
models inherently crafted for tweets, like \textsf{BTM} and \textsf{LLDA}. In
contrast, \textsf{HP} forms longer pseudo-documents, providing
more distinctive patterns to topic models, similar to \textsf{UP}. Its performance, though, is lower than \textsf{UP}, 
due to its insufficient coverage of documents: tweets with no hashtag are treated as
individual documents.

%\vspace{-12pt}
\section{Related Work}
\label{sec:relatedWork}

There has been a bulk of work on recommender systems over the years \cite{adomavicius2005toward,DBLP:books/sp/Aggarwal16}. 
Most recent works focus on microblogging platforms and Social Media, employing the bag model in order to
suggest new followees \cite{chen2009make}, URLs \cite{chen2010short} and
hashtags \cite{kywe2012recommending}. Others employ topic models for the same
tasks, e.g., hashtag recommendations \cite{godin2013using}. In the
latter case, emphasis is placed on tackling sparsity through \textit{pooling
techniques}, which aggregate short texts that share similar
content, or express similar ideas into lengthy pseudo-documents
\cite{alvarez2016topic,mehrotra2013improving}. E.g., Latent Dirichlet
Allocation 
%(\textsf{LDA}) 
\cite{blei2003latent} and the Author Topic
Model \cite{rosen2004author} are trained on individual messages and on
messages aggregated by user~and~hashtag~in~\cite{hong2010empirical}.

Content-based PMR has attracted lots of interest in the data management
community \cite{DBLP:journals/pvldb/0001LZC16,hierarchicalTopicModels,
DBLP:journals/pvldb/El-KishkySWVH14,
DBLP:journals/pvldb/YuCZS17,
DBLP:journals/pvldb/SharmaJBLL16}, where 
many works aim to improve the time efficiency of topic models. \cite{hierarchicalTopicModels} parallelizes the training of \textsf{HLDA} through a novel concurrent dynamic matrix and a distributed tree. \cite{DBLP:journals/pvldb/0001LZC16} scales up \textsf{LDA} through the \textsf{WarpLDA} algorithm, which achieves $O(1)$ time complexity per-token and fits the randomly accessed memory per-document in the L3 cache. Along with other state-of-the-art \textsf{LDA} samplers, this work is incorporated into \textsf{LDA*}, a high-end system that scales \textsf{LDA} training to voluminous datasets, using 
%real-world clusters and 
different samplers for various types of documents \cite{DBLP:journals/pvldb/YuCZS17}. Another approach for the massive parallelization of \textsf{LDA} is presented in \cite{DBLP:journals/pvldb/SmolaN10}. Other works facilitate real-time content recommendations in Twitter. This is done either by partitioning the social graph across a cluster in order to detect network motifs in parallel \cite{DBLP:journals/pvldb/GuptaSGGZLL14}, or by holding the entire graph in the main memory of a single server in order to accelerate random walk-based computations on a bipartite interaction graph between users and tweets~\cite{DBLP:journals/pvldb/SharmaJBLL16}.

On another line of research, external resources are employed in order to 
augment text representation and improve their performance in various
tasks. For short text clustering, Dual Latent Dirichlet Allocation learns topics
from both short texts and auxiliary documents \cite{jin2011transferring}.
For personalized Twitter stream filtering, tweets can be transformed into RDF
triples that describe their author, location and time in order to use
ontologies for building user profiles \cite{kapanipathi2011personalized}.
User profiles can also be enriched with Wikipedia concepts
\cite{lu2012twitter} and with concepts from news
articles that have been read by the user \cite{ijntema2010ontology}.
These approaches lie out of the scope of our analysis,
which focuses on recommendations based on Twitter's
internal content.

To the best of our knowledge, no prior work 
%experimental analysis 
examines systematically Content-based PMR with respect to the aforementioned parameters, considering a wide range of options for each one.
%compares most state-of-the-art representation models on text-based PMR. 
%In most cases, topic models are merely compared with few other
%methods on other text mining tasks, such as document classification \cite{yan2013biterm}.
%In fact, the works closest to ours are two studies that compare various representation sources in combination with the bag model for two different tasks: 
%For example, Abel et al.~\cite{abel2011analyzing} compare
%hashtag-based, entity-based and topic-based user models in news recommendation, whereas Twittomender \cite{hannon2010recommending}
%considers models built from a user's tweets, a user's followees' and followers'
%tweets and combinations of them, testing their effectiveness in recommending new
%followees. We go
%beyond both works by examining a wider set of performance aspects for 
%representation models.

\section{Conclusions}
\label{sec:conclusions}
%\vspace{-4pt}

We conclude with the following five observations about Content-based Personalized Microblog Recommendation:

\emph{(i)} The token-based vector space model achieves the best
balance between effectiveness and time efficiency. In most cases, it offers
the second most accurate recommendations, while involving the minimum time
requirements both for training a user model and applying it to a test set. On
the flip side, it involves four parameters (i.e., degrees of freedom), thus
being sensitive to its configuration.

\emph{(ii)} The token n-gram graphs achieve the best balance between
effectiveness and robustness. Due to the global contextual information they
capture, they consistently outperform all other representation models to a
significant extent, while exhibiting limited sensitivity to their configuration.
Yet, they are slower than the vector space model by an order of magnitude, on
average.

\emph{(iii)} The character-based models underperform their
token-based counterparts, as their similarity measures cannot tackle the noise
that piles up when assembling document models into user models.

\emph{(iv)} The topic models exhibit much lower effectiveness than
the token-based bag models for three reasons: \emph{1)} most of them are
not crafted for the sparse, noisy and multilingual content of Twitter, \emph{2)}
they depend heavily on their configuration, and \emph{3)} they are
context-agnostic, ignoring the sequence of words in documents. 
Their processing is time-consuming, due to the inference of topics,  requiring parallelization techniques to scale to voluminous data \cite{hierarchicalTopicModels,
DBLP:journals/pvldb/SmolaN10,DBLP:journals/pvldb/YuCZS17}.

\emph{(v)} All representation models perform best when they are
built from the retweets of hyperactive users (information producers).

In the future, we plan to expand our comparative analysis to other
recommendation tasks for microblogging platforms, such as followees and hashtag suggestions.
%We also intend to examine whether purely sequential models, such as HMMs,
%an provide added value. 
% Another interesting line of research would be to test
% whether facing the user recommendation problem as a regression problem, while
% keeping the examined representations, can significantly improve the performance
% of the systems. 

\balance

\bibliographystyle{abbrv}
\bibliography{bibliography}

\end{document}